\documentclass{aa}  

\usepackage{changepage}
\usepackage{graphicx}
\usepackage[dvipsnames]{xcolor}
\usepackage[colorlinks=true,allcolors=blue]{hyperref}
\usepackage[capitalize]{cleveref}
\usepackage{arydshln}
\usepackage{natbib}
\usepackage{adjustbox}
\usepackage{tcolorbox}
\usepackage{mathrsfs,amsmath} 
\usepackage{comment}
\usepackage{gensymb}
\usepackage{booktabs}
\usepackage{adjustbox}
\usepackage[normalem]{ulem}
\usepackage[varg]{txfonts}
\usepackage{caption}
\usepackage{subfig}
\usepackage{multirow}

\crefname{section}{Sect.}{Sects.}

\newcommand{\br}[0]{\mathbf{r}}
\newcommand{\bk}[0]{\mathbf{k}}
\newcommand{\bu}[0]{\mathbf{u}}

\begin{document} 

\title{Horizontal velocity eigenfunctions of solar inertial modes using local correlation tracking of magnetic features}

\author{
Neelanchal~Joshi \inst{\ref{mps}}
\and
Zhi-Chao~Liang \inst{\ref{mps}}
\and
Damien~Fournier \inst{\ref{mps}}
\and
Yash~Mandowara \inst{\ref{mps}}
\and 
Richard~S.~Bogart \inst{\ref{stanford}}
\and
Suprabha~Mukhopadhyay \inst{\ref{mps}}
\and
Laurent~Gizon \inst{\ref{mps},\ref{gottingen}}\thanks{Corresponding author: \email{gizon@mps.mpg.de}}
}

\institute{
Max-Planck-Institut f\"ur Sonnensystemforschung, Justus-von-Liebig-Weg 3, 37077 G\"ottingen, Germany \label{mps}
\and
W.\ W.\ Hansen Experimental Physics Laboratory, Stanford University, Stanford, CA 94305, USA \label{stanford}
\and
Institut f\"ur Astrophysik und Geophysik, Georg-August-Universit\"at G\"ottingen, Friedrich-Hund-Platz 1, 37077 G\"ottingen, Germany \label{gottingen}
}

\date{Received $\langle\ldots\rangle$ / Accepted $\langle\ldots\rangle$}
 
 \abstract
    {Solar inertial modes enable probes of the bulk of the convection zone over a broad latitude range. A proper study of high-latitude modes requires near-surface horizontal flow maps above $65\degree$ latitude at a cadence of at least one day. Previous observations have utilised ring-diagram analysis (RDA) of acoustic modes and local correlation tracking (LCT) of granules, but these methods do not have the capacity to access higher latitudes without compromising the signal-to-noise ratio (S/N).}
    {We measured flows associated with solar inertial modes using local correlation tracking of small magnetic features in SDO/HMI magnetograms separated by 1~h (LCTMag). This non-helioseismic method extracts horizontal near-surface flows with a good S/N up to $85\degree$ latitude. To carry out a comparison below $65\degree$, we also measured flows using LCT of granulation features in continuum intensity images separated by 45~s (LCTGran).}
    {We constructed longitude-latitude maps of horizontal flows at a cadence of 6~h from LCTMag and LCTGran using overlapping $5\degree$ tiles. We additionally applied the RDA pipeline to HMI Dopplergrams over the same period to obtain flow maps at a cadence of 9~h. After removing large-scale systematics and long-term variations, we applied spectral analysis to identify modes in the inertial frequency range and extracted their spatial shapes via singular value decomposition.}
    {We find that the noise background is approximately 80\% less in the power spectrum of LCTMag than in the power spectrum of LCTGran. Mode shapes are consistent between LCTMag and LCTGran below $65\degree$, but granulation tracking performs poorly at higher latitudes due to insufficient granule resolution. LCTMag provides higher-latitude access than both RDA and LCTGran with better S/N values, resulting in reduced spectral leakage from the window function. We also report a new retrograde inertial mode in the LCTMag dataset at frequency $-192$~nHz in the Carrington frame with a longitudinal wavenumber of $m=3$. This mode is seen at all latitudes; its radial vorticity is symmetric across the equator, with a pronounced spiral pattern above $65\degree$ and a maximum velocity amplitude of 1~m/s near $70\degree$. Finally, we propose a possible mode identification using a linear eigenvalue solver.}
    {}

   \authorrunning{Joshi et. al.}
   \titlerunning{Horizontal velocity eigenfunctions of solar inertial modes using
local correlation tracking of magnetic features}
   \maketitle

\section{Introduction \label{sec:intro}}
The Sun exhibits a rich spectrum of global oscillations that serve as probes of its interior dynamics and structure. Classical helioseismology relies predominantly on acoustic modes and surface gravity modes, which are used to infer the structure, flows, and rotation of the solar interior \citep[e.g.,][]{jcd2002,burston2015}. The quasi-toroidal low-frequency inertial modes offer unique diagnostic potential, being sensitive to differential rotation, superadiabaticity, turbulent viscosity and entropy gradients in the deep convection zone \citep{Gizon2021, Hanson2022, bekki2024, Gizon2024, dey2026}. Restored by the Coriolis force, these modes have characteristic periods of the order of the solar rotation period ($\sim$27 days). Observational detection of inertial modes has been made possible by the Helioseismic and Magnetic Imager aboard the Solar Dynamics Observatory \citep[SDO/HMI;][]{Scherrer2012, Schou2012}, leading to the identification of several distinct families: equatorial Rossby modes \citep{Loeptien2018}, critical-layer modes, high-latitude modes \citep{Gizon2021}, and high-frequency retrograde (HFR) modes \citep{Hanson2022}.

Contemporary observational studies of inertial modes have employed two broad classes of measurement techniques. Helioseismic methods, including Ring-Diagram Analysis (RDA; \citealt{Proxauf2020, Hanson2020, Gizon2021}), time-distance helioseismology \citep{liang2019, ding2025}, and normal-mode coupling analysis \citep{hanasoge2019,mandal2020} have been successful in classifying the spectrum of observed modes and extracting their latitudinal profiles. Non-helioseismic approaches, such as local correlation tracking (LCT) of granules, complement these by measuring proper horizontal photospheric motions directly \citep{Loeptien2018}. First introduced by \cite{november&simon1988}, LCT tracks surface features by measuring the average displacement of structures within a region of interest, and has been widely used since. LCT has allowed the first detection of Rossby modes on the Sun \citep{Loeptien2018}. Even though these methods have proven successful in detecting weak inertial mode signals, they often face challenges accessing high latitudes. 

Cross-correlation techniques applied to photospheric magnetograms have previously been used to measure large-scale solar flows, including differential rotation and meridional circulation \citep{meunier1999,meunier2005,hathaway2011,rightmire2012measurements}. In this paper, we apply LCT to magnetic network features in HMI magnetograms (hereafter, LCTMag) to measure horizontal flows associated with solar inertial modes. Compared with RDA and LCT of granules (hereafter LCTGran), LCTMag provides more complete spatial coverage and extends reliable measurements to higher latitudes. We employed both LCTMag and LCTGran to measure solar surface flows using 14 years of HMI magnetogram and continuum intensity data, respectively, from which horizontal velocity maps are derived (\cref{sec:hflows}). We analyse these flow maps to extract the power (\cref{sec:im_ident}) and latitudinal profiles of inertial modes (\cref{sec:ef_extract}), with closer attention paid to several example modes. These maps are further decomposed into radial vorticity and horizontal divergence to identify HFR modes, and the mode power and latitudinal profiles are compared against RDA (\cref{sec:rda_compare}). We also report the identification of a new $m=3$ mode not previously reported in the literature (\cref{sec:m3_mode}). Our results provide fresh observational constraints for theoretical models and numerical simulations, and the horizontal flow measurements serve as a complementary dataset for the independent verification of known modes and the potential discovery of new ones (\cref{sec:conclusions}). Preliminary results for high-latitude inertial mode eigenfunctions using LCTMag were presented by \cite{joshi2025}.

\section{Measuring surface horizontal flows \label{sec:hflows}} 

\subsection{Datasets}

For our LCT analysis, we retrieved the full-disk images from the \texttt{hmi.m\_720s} and the \texttt{hmi.ic\_45s} data series from the Joint Science Operations Center\footnote{\url{http://jsoc.stanford.edu/}}. The \texttt{hmi.m\_720s} series contains line-of-sight (LOS) magnetograms at a 12~min cadence, while the \texttt{hmi.ic\_45s} series provides continuum intensity images at a higher cadence of 45~s. These images are used to infer horizontal flows using passive tracers such as magnetic network elements and granules. The time range used for this analysis is May 2010 to December 2024.

Complementing the LCT data, we used the $5\degree$-tile data obtained from the HMI RDA pipeline \citep{Bogart2011a, Bogart2011b}. The $5\degree$ tiles are used instead of the $15\degree$ tiles for better spatial resolution, utilising a $2.5\degree$ sampling grid. For a proper comparison, we used the same grid sampling and tile size for LCT measurements. Since the RDA pipeline provides flow maps approximately every 9~h, we selected the RDA data from May 1, 2010 to December 31, 2024, closely matching the LCT dataset.

\subsection{Inferring horizontal flows using local correlation tracking of granules and magnetic features}

For a given target location $(\theta, \phi)$ on the solar surface, where $\theta$ denotes the co-latitude and $\phi$ the heliographic longitude, we consider a $5\degree\times 5\degree$ spatial patch centred at this point. This patch is imaged at two distinct times $t$ and $t+\delta t$. The time separation $\delta t$ is chosen such that the evolution of the features during $\delta t$ is negligible. For granulation, $\delta t$ is taken to be 45~s (the typical lifetime of granules is about 10~min), while for magnetic features, $\delta t$ is 1~h. The patches are tracked at the Carrington rotation rate of $-456$~nHz with $t$ as the reference time to remove the apparent motion due to solar rotation, and then remapped using Postel's azimuthal equidistant projection, with the target location $(\theta, \phi)$ as the projection centre. The images were apodized using a Tukey window consisting of a flat central region with a radius of about $1\degree$, surrounded by raised-cosine tapers toward the edges, to reduce edge effects. LCT is known to suffer from an underestimation bias, which has been attributed to the Gaussian apodization function being centred at the same pixel in both images, even though the features have shifted \citep{fisher2020}. Since the flat-top region of the Tukey window provides nearly uniform weighting over the central region, this underestimation bias is negligible in our horizontal flow maps.

\begin{figure}
    \centering
    \begin{minipage}[c]{0.49\linewidth}
        \centering
        \includegraphics[width=\linewidth]{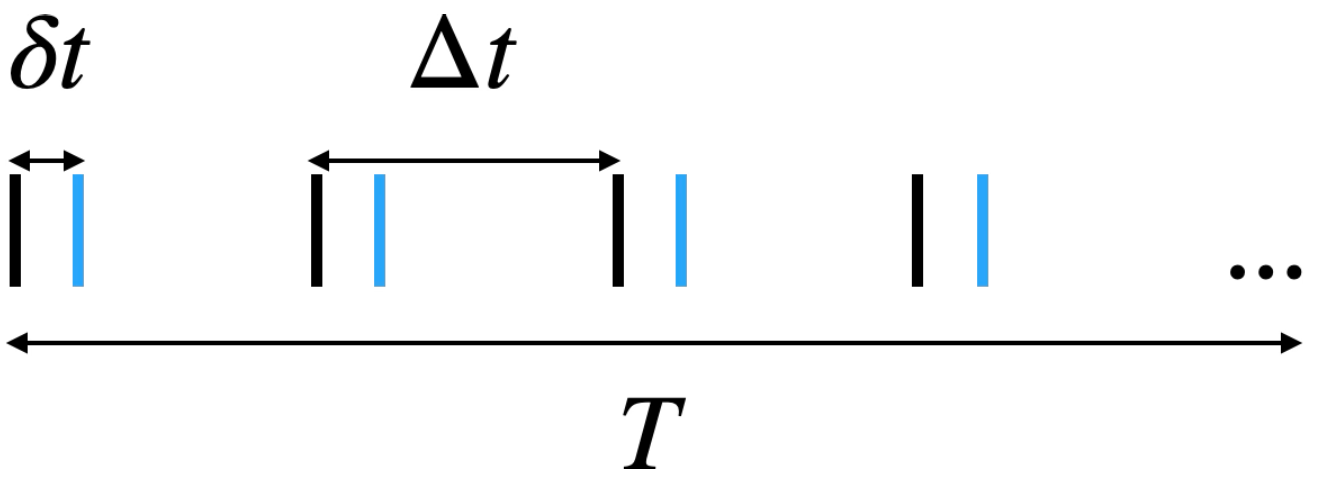}
    \end{minipage}
    \hfill
    \begin{minipage}[c]{0.49\linewidth}
        \centering
        \begin{tabular}{lcc}
            \hline\hline\noalign{\vskip 1pt}
            & LCTMag & LCTGran \\
            \hline
            $\delta t$ & 1~h & 45~s \\
            $\Delta t$ & 2~h & 30~min \\
            $T$        & 6~h & 6~h \\
            \hline
        \end{tabular}
    \end{minipage}
    \caption{Schematic (\textit{left}) alongside the values (\textit{right}) for the various parameters used in the flow measurements through LCTMag and LCTGran. The image pairs used for cross-correlation are separated by $\delta t$. The pairs are separated by $\Delta t$ and the CCFs are averaged over $T$.}
    \label{fig:schematic}
\end{figure}

For a $5\degree \times 5\degree$ patch, the metric distortion of the 
Postel projection is negligible and the projected coordinates $\br=(x, y)$ 
can be treated as locally Cartesian. For each target location $(\theta, \phi)$, we compute the spatial cross-correlation between the LOS magnetogram patches $B(\br,t)$ and $B(\br, t + \delta t)$ via the discrete Fourier transform,
\begin{equation}
C(\delta \mathbf{r}, t) = \sum_{\bk}
B^*(\bk, t)\, B(\bk, t+\delta t)\, 
\exp(-\mathrm{i}  \bk \cdot \delta \br),
\end{equation}
where
\begin{itemize}
    \item $\delta \mathbf{r} = (\delta x, \delta y)$ is the spatial shift between the two images;
    \item $B(\bk, t)$ is the discrete spatial Fourier transform of the magnetogram patch at time, $t$;
    \item $\bk =(k_x, k_y)$ is the horizontal wavevector;
    \item The superscript $^*$ denotes complex conjugation. 
\end{itemize}
For a local correlation tracking of granulation (LCTGran), the magnetograms, $B$, are replaced by continuum intensity images, $I$.

\begin{figure*}
\sidecaption
  \includegraphics[width=12cm]{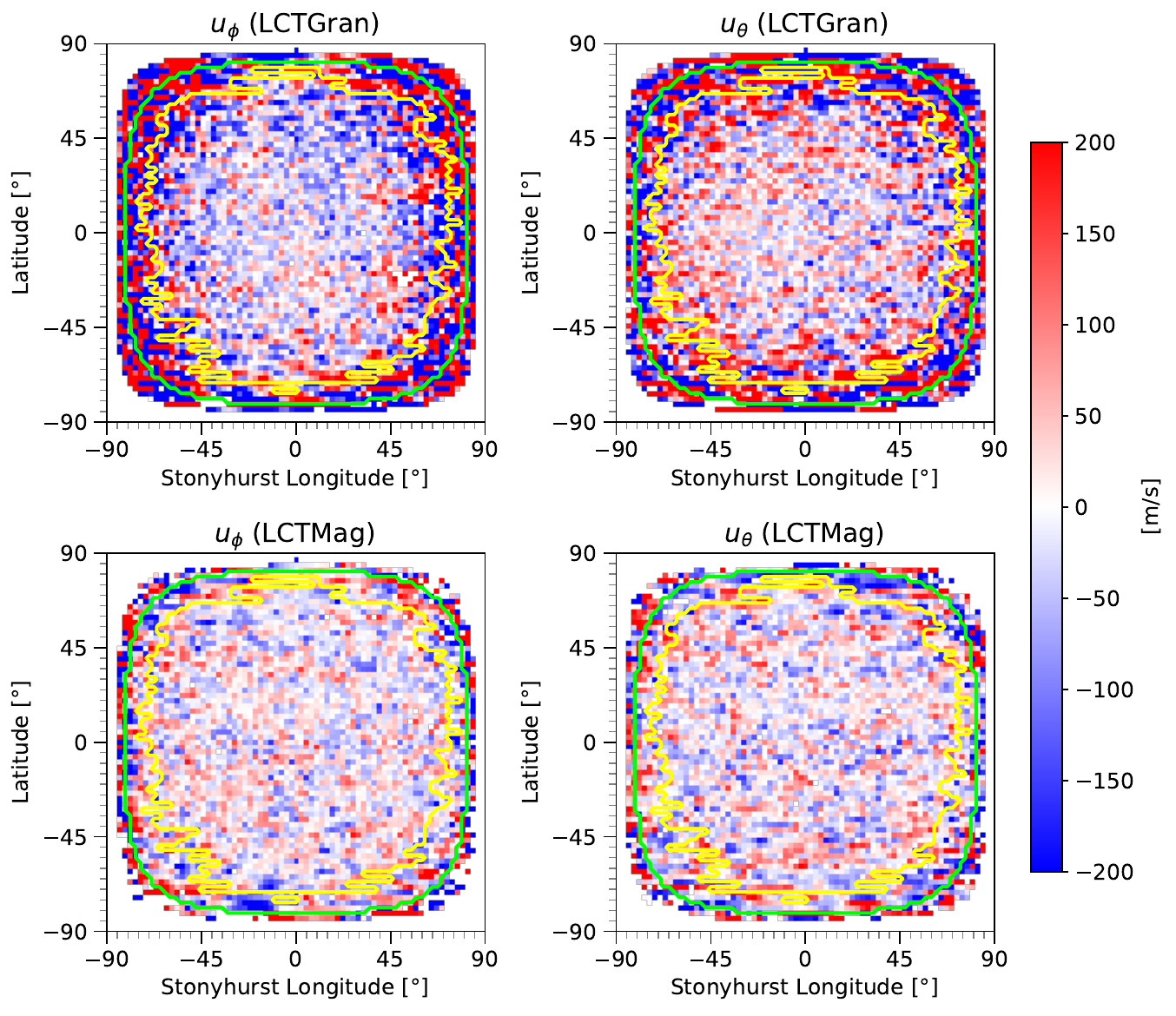}
  \caption{  \label{fig:flow}
  Example horizontal flow maps derived from granulation (\textit{top}) and magnetic feature tracking (\textit{bottom}) using HMI data from 2011-06-07 00:00--06:00 TAI. The lime contour marks the adopted cropping extent ($0.99$ of the solar disk radius), while the yellow contour indicates the boundary of the RDA spatial coverage for the closest time period (2011-06-06 22:35:37 to 2011-06-07 07:40:07 TAI).
  }
\end{figure*}

After obtaining $C(\delta \br, t)$, we averaged multiple realisations of the cross-correlation function (CCF) over a time interval, $T$, to reduce random noise and obtain a smooth CCF profile. \Cref{fig:schematic} shows a schematic and the values of the various temporal parameters used to compute the horizontal flows. Each realisation is separated by the next one in time, represented by $\Delta t$. For granules, this time is chosen to be 30~min, while for magnetograms, this is chosen to be 2~h. The CCF realisations are averaged over a time interval of $T = 6~\text{h}$ to produce the final CCF. \Cref{fig:ccf_2d} shows an example CCF at $(\theta, \phi) = (60\degree, 30\degree)$ obtained from LCTGran and LCTMag. The left column shows the CCF without averaging, while the right column shows the CCF after averaging over 6~h. We can clearly see that the averaged CCFs have a much smoother profile and a much sharper central peak, leading to more accurate peak localisation.

We measured the shift of the peak of the mean cross-correlation function from the centre along the direction of interest, namely, along $x$ and $y$ in the local Cartesian coordinates. We determined the shifts by fitting a 2D quadratic surface to the CCF around its maximum. To avoid the phenomenon of peak-locking \citep[e.g.,][]{Langfellner2015,mahajan2021}, we iteratively shifted the CCF by $-\delta x$ and $-\delta y$. The peak of this shifted CCF is measured again using the 2D surface fit and the procedure was iterated three times. The $\delta x$ and $\delta y$ calculated at each step are added together to get the total shift. The total shifts $\sum \delta x$ and $\sum \delta y$ are divided by $\delta t$ to obtain the final horizontal flow components $u_x$ and $u_y$. These horizontal flow components $(u_x, u_y)$ are equivalent to the standard data products $(u_\phi, -u_\theta)$, with a sign change for the $\theta$ component (since it is positive in the southward direction by convention).

We quantified the spatial extent of LCTGran using the width of the CCF near its maximum. The full width at half maximum (FWHM), measured directly from cuts along the $x$- and $y$-directions, is obtained as the distance between points where the CCF falls to half its peak value (with interpolation where required). \Cref{fig:ccf_widths_compare} shows the FWHM averaged over a 6~h period at each disk position for $B_0 \approx 0$, where $B_0$ is the heliographic latitude of the disk center. LCTMag tracks features with widths of 2--5~Mm, consistent with inter-supergranular magnetic network scales, whereas LCTGran is limited to smaller spatial extents and degrades toward the limb, with a sharp FWHM increase beyond $\sim65^\circ$ latitude. In contrast, LCTMag shows decreasing feature size at higher latitudes, likely due to projection and foreshortening. The widths along the east-west limb show an increase in both LCTGran and LCTMag, which could be attributed to the apparent shift of the features due to tracking.

\begin{figure*}[h!]
\sidecaption
  \includegraphics[width=12cm]{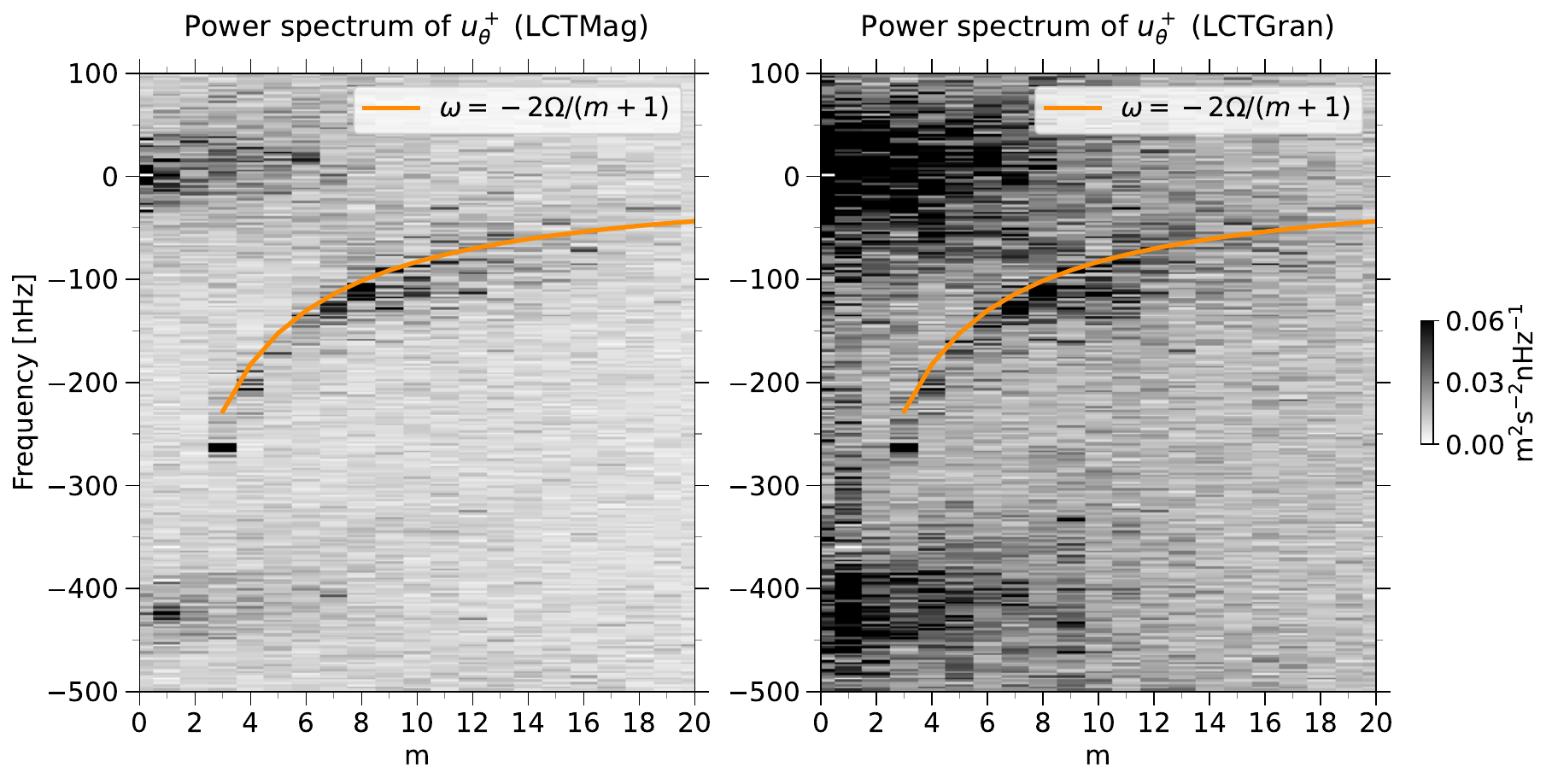}
  \caption{  \label{fig:pow2d_rossby}
  Power spectra of $u_\theta^+$ from magnetic feature tracking (LCTMag; \textit{left}) and granulation tracking (LCTGran; \textit{right}) as a function of frequency and azimuthal wavenumber $m$, computed over the period 2010--2024. The power spectra are averaged over latitudes within $\pm30\degree$. The orange curve shows the theoretical dispersion relation for sectoral Rossby waves, $\omega = -2\Omega_\text{Carr}/(m+1)$, where $\Omega_\text{Carr}$ is the Carrington rotation rate. The curve is shown only for $m\geq3$, corresponding to the range of observed modes.
  }
\end{figure*}

\subsection{Post-processing of horizontal flow maps}

Horizontal flow maps derived from correlation tracking contain systematic errors that dominate the power spectrum. One of the most significant is the shrinking-Sun effect \citep{loeptien2016_shrinkingsun}: a spurious large-scale gradient toward the disk centre in the flow maps obtained from intensity images, primarily attributed to insufficient resolution of solar granules. Flow maps inferred from magnetic network tracking also show large-scale gradients, but directed away from the disk centre \citep{mahajan2021}.

We first subtract the long-term median at each grid point from the flow maps in the Stonyhurst coordinate system. This removes the large-scale gradients such as the shrinking Sun effect. We classified flow values that are higher than 3 $\sigma$ from the time series at each grid point as outliers and mask them.

The annual variation of the $B_0$ angle introduces systematic yearly variations in the flow maps. This also prevents us from having a full view of both the north and south poles of the Sun throughout the year. To mitigate these effects, we fit sinusoidal functions with periods of one year and six months to each time series and subtracted the fitted variations.

\Cref{fig:flow} shows the cleaned horizontal flow maps derived from granulation and magnetic feature tracking. We crop the flow maps at $0.99$ of the solar disk radius, which provides an optimal balance between suppressing limb noise and retaining high-latitude signal. We overlay the contours of the 0.99 radius and the boundary of the RDA flow map from the closest time period. LCT provides broader spatial coverage and access to higher latitudes. Because LCTGran and LCTMag are sensitive to different flow scales, their maps differ even for the same time period, with LCTGran showing higher noise levels.

From these horizontal flow maps, we also calculated the radial vorticity, $\zeta_\mathrm{r}$, and horizontal divergence, $\nabla \cdot \mathbf{u}_h$, where $\mathbf{u}_h = (u_\theta, u_\phi)$. We used the Python package \texttt{shtns} \citep{shtns} to project each flow map onto a basis of spherical harmonics. This approach allows for spatial derivatives to be computed spectrally rather than by using finite-difference schemes, thereby avoiding numerical noise and grid-scale artefacts associated with central differencing. As a result, the derived vorticity and divergence fields are significantly smoother and more robust, particularly at small spatial scales.

\begin{figure*}[h!]
\sidecaption
  \includegraphics[width=12cm]{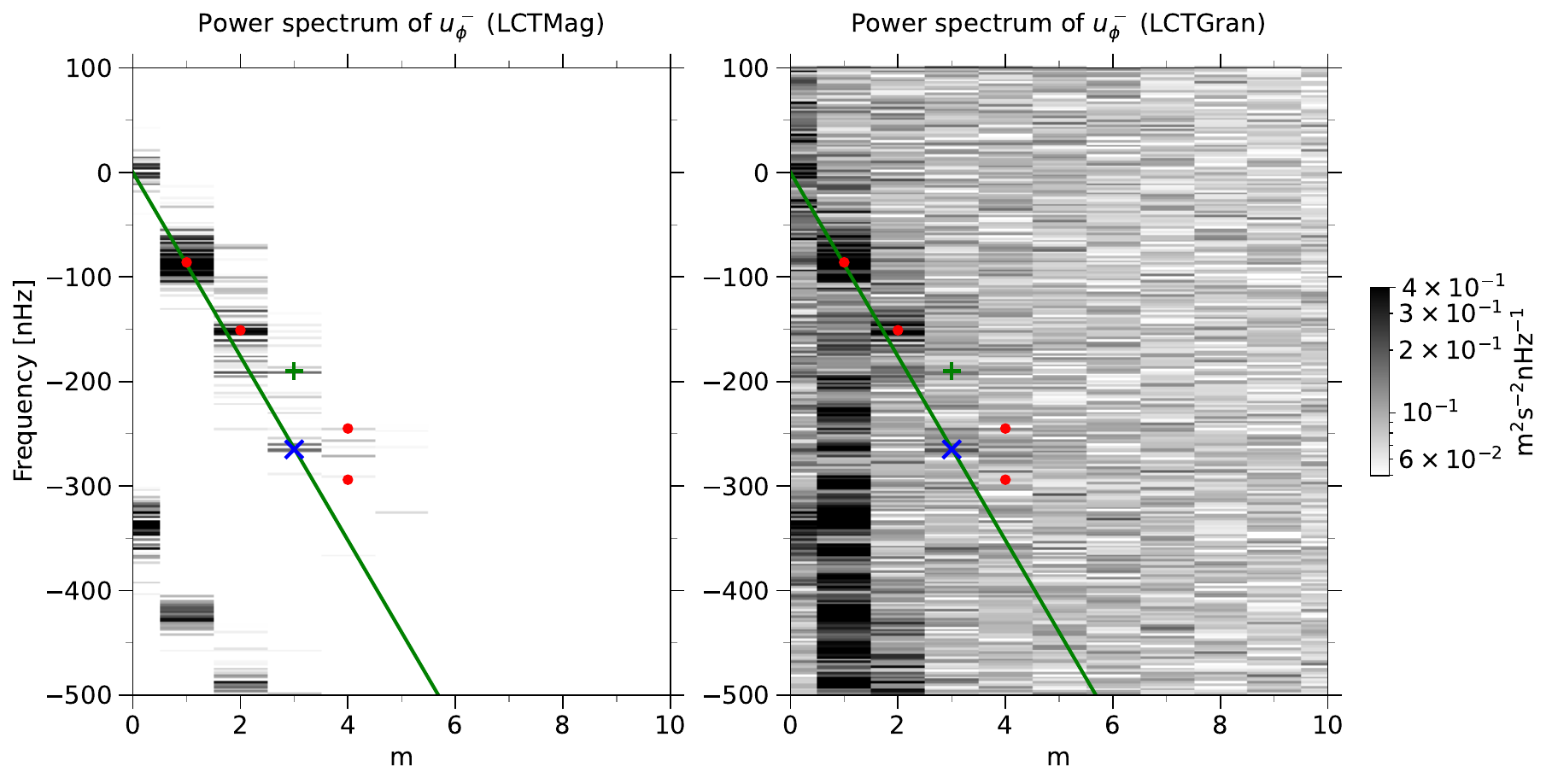}
  \caption{ {\label{fig:pow2d_HL}}
  Power spectra of $u_\phi^-$ from magnetic feature tracking (LCTMag; \textit{left}) and granulation tracking (LCTGran; \textit{right}) as a function of frequency and azimuthal wavenumber $m$ for the time period 2010--2024. The power spectra are averaged over latitudes between $45\degree$ and $75\degree$. The power is displayed on a logarithmic scale for clarity. The peak frequencies of the high-latitude modes reported by \cite{Gizon2021} are shown as red points. The dark green straight line is drawn through zero frequency at $m=0$ and the $m=1$ mode frequency. The blue cross indicates the power from the $m=3$ equatorial Rossby mode at high latitude, while the green plus marks the newly discovered $m=3$ mode (see \cref{sec:m3_mode}).}
\end{figure*}

\begin{figure*}[h!]
\sidecaption
  \includegraphics[width=12cm]{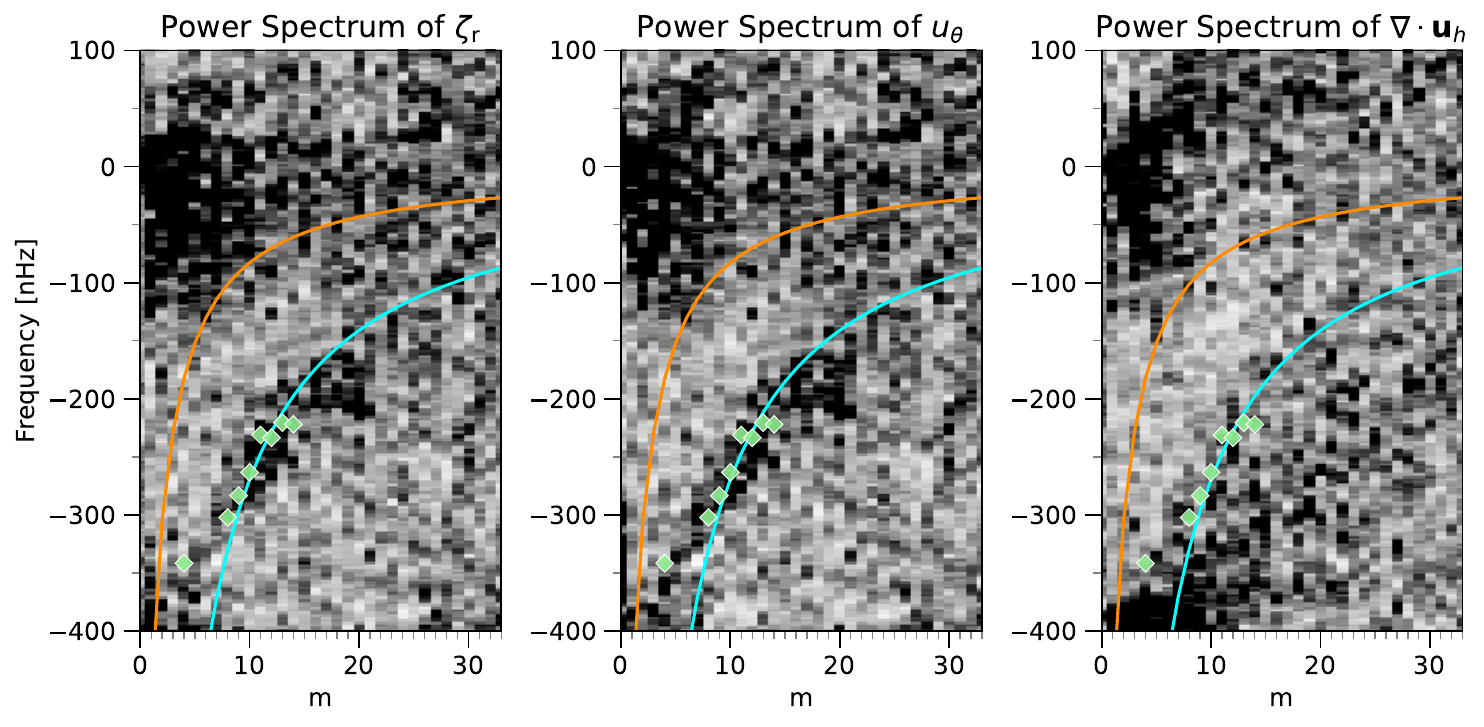}
  \caption{  \label{fig:powhfr}Normalised power spectra of $\zeta_\mathrm{r}$ (\textit{left}), $u_\theta$ (\textit{middle}), and $\nabla \cdot \mathbf{u}_h$ (\textit{right}) obtained from magnetic feature tracking (LCTMag) for the spherical harmonics $l=m+1$ (for $\zeta_\mathrm{r}$ and $u_\theta$) and $l=m$ (for $\nabla \cdot \mathbf{u}_h$), computed over the period 2010--2024. The orange ridge corresponds to the classical sectoral Rossby modes. The cyan line indicates the approximate frequency of the HFR modes, about three times that of the sectoral Rossby modes. Each power spectrum is smoothed by convolving with a window of 12.5~nHz. The light green dots are the mode frequencies reported in \cite{Hanson2022}. 
  }
\end{figure*}

\section{Identifying inertial modes on the solar surface \label{sec:im_ident}}

To successfully identify the inertial mode peaks, we first need to transform our horizontal flow components $(u_{\phi}(\theta, \phi, t), u_{\theta}(\theta, \phi, t))$ to the spectral domain. The cleaned flows are converted from the Stonyhurst frame to a co-rotating frame of reference with the Carrington rate of rotation and separated into their symmetric and anti-symmetric components with respect to the equator by
\begin{equation}
    u^{\pm}_j(t,\theta,\phi)
    =
    \frac{1}{2}
    \left[
    u_j(t,\theta,\phi)
    \pm
    u_j(t,\pi-\theta,\phi)
    \right],
\end{equation}
where $j$ denotes the component of the horizontal flow (either $\theta$ or $\phi$). The symmetric part of the flow component is given by $+$ and the anti-symmetric component by $-$. The flows are then transformed to the spectral domain by a 2D discrete Fourier transform in time and longitude,
\begin{equation}
U^{\pm}_j(\omega, \theta, m) = \frac{1}{\eta(\theta)\,n_t n_\phi} 
\sum_{t}\sum_{\phi} 
u^{\pm}_j(t, \theta, \phi)\,
\exp\!\left[\mathrm{i}(m\phi - \omega t)\right].
\end{equation}
The quantities $\omega$, $m$, $\eta(\theta)$, $n_t$, and $n_\phi$ are defined as:
\begin{itemize}
    \item $m$ is the longitudinal wavenumber and $\omega$ is the angular frequency measured in the Carrington frame. We restricted $m$ to positive integers, so that negative (positive) values of $\omega$ correspond to retrograde (prograde) propagation;
    \item $\eta(\theta)$ is the fraction of available data at colatitude $\theta$, used to correct for incomplete temporal or longitudinal coverage;
    \item $n_t$ is the total number of time samples in the dataset;
    \item $n_\phi=144$ is the total number of longitude grid points.
\end{itemize}
The power spectral density (PSD) of the flow component is then given by
\begin{equation}
P^{\pm}_j(\omega, \theta, m) = \left| U^{\pm}_j(\omega, \theta, m) \right|^2,
\end{equation}
and is subsequently normalized by the frequency-bin width ($\sim$2~nHz) to give units of m$^2$/s$^2$/nHz.

\subsection{Dispersion relations seen in 2D power spectra in the inertial frequency range}

\Cref{fig:pow2d_rossby} shows the PSD computed from the $u^+_\theta$ component of the flow averaged between $\pm 30\degree$ latitude, for granulation and magnetic feature tracking. The theoretical sectoral Rossby mode dispersion relation is overplotted. This ridge is more clearly visible in the $l=m$ components of radial vorticity as seen in \cref{fig:pow2d_rvort_mag_gran}. We can clearly see the equatorial Rossby mode ridge in the data. The region close to zero-frequency in \cref{fig:pow2d_rossby} shows the convection noise, which is much higher in granulation tracking. Due to the window function, the convection signal near zero frequency produces sidelobes separated by the synodic Carrington rotation frequency ($424.3$~nHz). Within the frequency range shown, only the first sidelobe is visible.

\Cref{fig:pow2d_HL} shows the high-latitude (HL) modes in the $u_\phi^-$ PSD for LCTMag and LCTGran. This figure highlights the advantage of using magnetic network features as passive tracers. We can clearly see that at these latitudes, LCTGran shows considerable noise because granules cannot be reliably resolved, making the HL modes difficult to detect, whereas the HL modes are clearly detected in LCTMag. The colour scale has been saturated to highlight the influence of noise in the two datasets. The peak frequencies of the HL modes identified by \cite{Gizon2021} in the $u_\phi^-$ and $u_\theta^+$ components are marked by red dots. Power associated with the equatorial $m=3$ Rossby mode is also visible at high latitudes (blue cross). We additionally identified a new $m=3$ mode (green plus; see \cref{sec:m3_mode}).

\begin{figure*}
  \resizebox{\hsize}{!}{\includegraphics{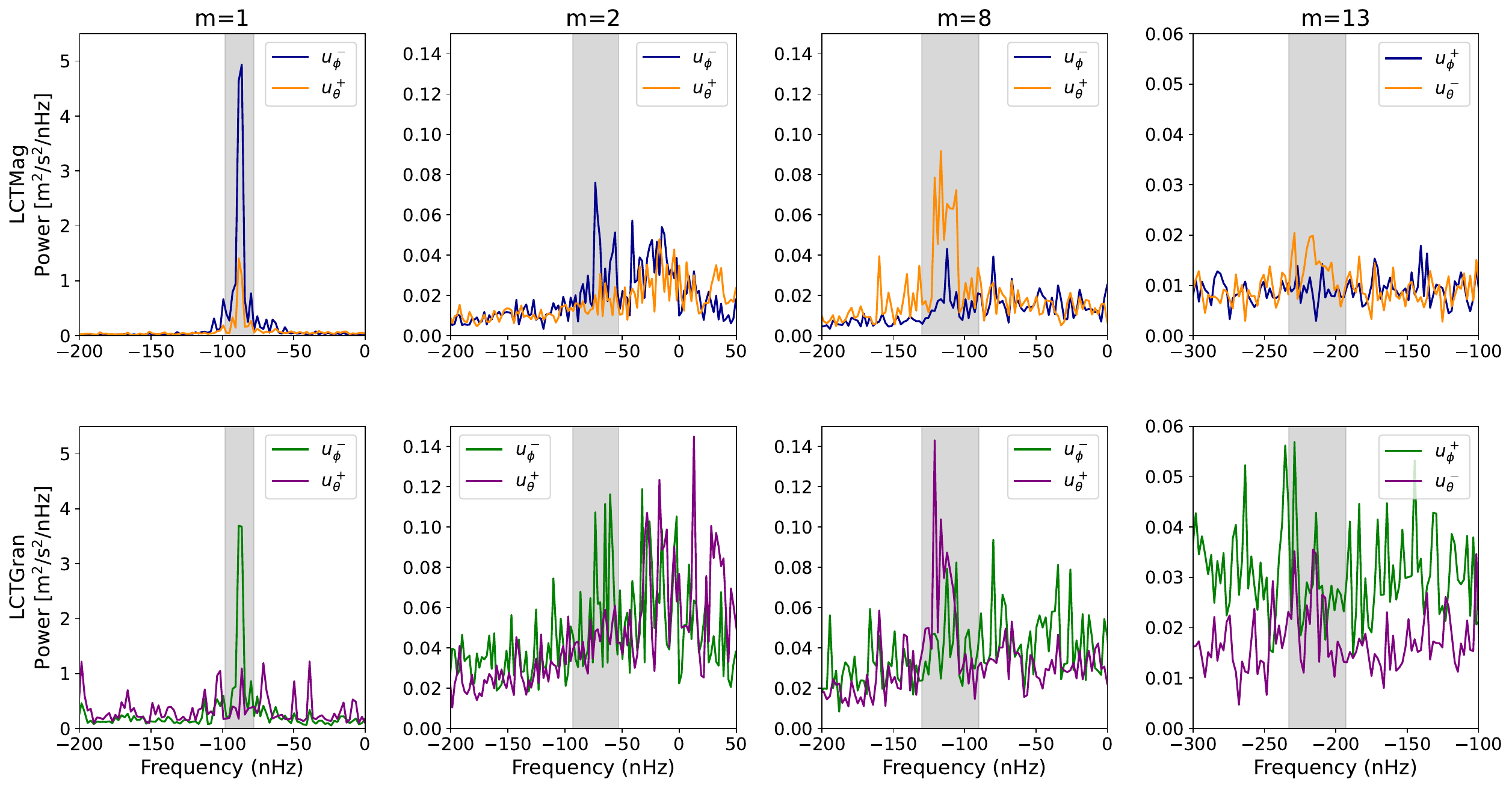}}
  \caption{  \label{fig:pow1d}
  $u_\theta$ and $u_\phi$ 1D power spectra of a few example inertial modes using LCTMag and LCTGran. The spectra are computed by averaging the power at specific latitude bands (see \cref{tab:lorentzian_fits}). The panels from left to right are the $m=1~(+)$ high-latitude mode, the $m=2~(+)$ critical-layer mode, the $m=8~(+)$ equatorial Rossby mode, and the $m=13~(-)$ high-frequency retrograde mode. The symmetry of the mode is given in brackets. The shaded gray regions indicate the frequency window used to isolate the mode in \cref{sec:ef_extract}. 
  }
\end{figure*}

For the high-frequency retrograde (HFR) modes, we analyzed the $l=m+1$ components of $\zeta_\mathrm{r}$ and $u_\theta$, following \cite{Hanson2022}. This ridge was previously reported only in $\zeta_\mathrm{r}$; however, the LCTMag dataset also reveals a clear detection in $u_\theta$ (see \cref{fig:powhfr}, left two panels). The LCTGran dataset also shows a region of excess power associated with these modes, although the signal is substantially weaker (see \cref{fig:pow2d_rvort_hfr}). The excess power close to zero frequency is due to convection noise.
\cite{hanson2025} report the radial velocity component of the HFR modes in the $l=m$ components of the horizontal divergence, $\nabla \cdot \mathbf{u}_h$. The right panel of \cref{fig:powhfr} shows the power spectra of the $l=m$ components of $\nabla \cdot \mathbf{u}_h$. A modest excess of power is visible along the HFR ridge, highlighted by the cyan curve, although the signal remains tenuous.

\begin{table*}
\centering
\caption{Measured parameters of global inertial modes.}
\label{tab:lorentzian_fits}
\resizebox{0.85\textwidth}{!}{
\begin{tabular}{lllcccc}
\hline\hline\noalign{\vskip 1pt}
Mode & Latitude band& Methods & $S$ & $\omega_0/2\pi$ & $\Gamma/2\pi$ & S/N\\
 &  &  & $(\rm m^2/s^2/nHz)$ & (nHz) & (nHz) &  \\
\hline\noalign{\vskip 2pt}
\multirow{3}{*}{\shortstack[l]{$m=1$, $u_\phi^-$  High latitude}}& \multirow{3}{*}{$45^\circ-75^\circ$}
 & LCTMag & $3.4^{+0.9}_{-0.8}$ & $-87.2^{+0.5}_{-0.5}$ & $7.2^{+1.2}_{-1.0}$ & $194.6$ \\[3pt]
 &  & LCTGran & $2.2^{+0.6}_{-0.5}$ & $-87.4^{+0.7}_{-0.7}$ & $9.8^{+1.9}_{-1.6}$ & $19.9$ \\[3pt]
 &  & RDA & $2.3^{+0.7}_{-0.6}$ & $-87.3^{+0.5}_{-0.5}$ & $7.1^{+1.3}_{-1.1}$ & $70.9$ \\[2pt]
\hline\noalign{\vskip 2pt}
\multirow{3}{*}{\shortstack[l]{$m=2$, $u_\phi^-$ \\ Critical Layer}}
 & \multirow{3}{*}{$15^\circ-45^\circ$}
 & LCTMag & $0.029^{+0.006}_{-0.004}$ & $-66.2^{+3.9}_{-3.1}$ & $37.0^{+9.5}_{-8.5}$ & $3.8$ \\[3pt]
 &  & LCTGran & $0.042^{+0.025}_{-0.007}$ & $-64.0^{+5.2}_{-3.7}$ & $26.1^{+11.8}_{-12.8}$ & $1.2^{\dagger}$ \\[3pt]
 &  & RDA & $0.026^{+0.005}_{-0.003}$ & $-65.4^{+3.1}_{-2.6}$ & $33.3^{+7.0}_{-6.6}$ & $6.1$ \\[2pt]
\hline\noalign{\vskip 2pt}
\multirow{3}{*}{\shortstack[l]{$m=8$, $u_\theta^+$ \\ Equatorial Rossby}}
 & \multirow{3}{*}{$0^\circ-30^\circ$}
 & LCTMag & $0.050^{+0.010}_{-0.008}$ & $-111.8^{+1.7}_{-1.7}$ & $26.4^{+5.1}_{-4.6}$ & $5.3$ \\[3pt]
 &  & LCTGran & $0.061^{+0.015}_{-0.010}$ & $-115.0^{+2.0}_{-2.0}$ & $27.6^{+6.3}_{-6.1}$ & $3.3$ \\[3pt]
 &  & RDA & $0.047^{+0.012}_{-0.009}$ & $-113.0^{+1.3}_{-1.3}$ & $18.0^{+3.7}_{-3.4}$ & $6.1$ \\[2pt]
\hline\noalign{\vskip 2pt}
\multirow{3}{*}{\shortstack[l]{$m=13$, $u_\theta^-$  High-frequency  retrograde}}& \multirow{3}{*}{$0^\circ-30^\circ$}
 & LCTMag & $0.009^{+0.005}_{-0.001}$ & $-217.1^{+3.6}_{-3.7}$ & $24.7^{+14.1}_{-12.4}$ & $1.1^{\dagger}$ \\[3pt]
 &  & LCTGran & $0.010^{+0.009}_{-0.001}$ & $-221.6^{+5.2}_{-5.3}$ & $30.4^{+27.8}_{-19.9}$ & $0.8^{\dagger}$ \\[3pt]
 &  & RDA & $0.006^{+0.004}_{-0.001}$ & $-219.7^{+3.9}_{-3.9}$ & $26.2^{+16.3}_{-14.0}$ & $1.0^{\dagger}$ \\[2pt]
\hline
\end{tabular}
}
\tablefoot{
For each mode, the parameters were measured by fitting a Lorentzian profile, $L(\omega) = \dfrac{S\,(\Gamma/2)^2}{(\omega - \omega_0)^2 + (\Gamma/2)^2} + N$, to each power spectrum averaged over a selected latitudinal band. Here, $S$ is the peak amplitude above the background noise level $N$, $\omega_0$ is the mode frequency, and $\Gamma$ is the FWHM.
Quoted uncertainties are $68\%$ confidence intervals estimated from Monte Carlo simulations.
$^\dagger$ At a low resolution of $\mathrm{S/N}\lesssim2$); the fitted parameters are not as well constrained.
}
\end{table*}

\cite{hanasoge2026} report prograde-propagating thermal Rossby modes below $0.98\,R_\odot$ using normal-mode coupling. \Cref{fig:pow2d_hanasoge} shows the radial vorticity power spectra derived from LCTMag data for the $l=m+1$ and $l=m$ components. We did not detect the prograde thermal Rossby ridge or the frequency splitting reported by \cite{hanasoge2026evidence}. Although magnetic features are thought to be anchored to deeper layers and advected by the flows there \citep{mahajan2021}, potentially providing LCTMag with sensitivity to those depths, we found no clear evidence of these modes in our data.

\subsection{Line profiles of the inertial modes}

At a particular azimuthal wavenumber, the averaged PSD $\langle P(\omega, \theta, m) \rangle_\theta$ gives us the line profiles for inertial modes at different latitude bands. For our analysis here, we choose an example mode from the four categories of inertial modes described in this text, as seen in \cref{fig:pow1d}, namely the $m=1$ HL mode, $m=2$ critical-layer mode, $m=8$ equatorial Rossby mode, and the $m=13$ HFR mode. We show the line profiles obtained from our LCT analysis using both granules and magnetic network features as tracers. Following the convention from \cite{charbonneau1999}, the modes symmetric in $u_\theta$ and antisymmetric in $u_\phi$ with respect to the equator are labelled as symmetric modes ($+$), while the opposite symmetry modes are labelled antisymmetric modes ($-$).

These line profiles show peaks of power associated with inertial modes at different $m$. We observe that the power is not the same for the different flow components, but is largely consistent in terms of the peak frequency. LCTGran exhibits much higher noise than LCTMag, especially at higher latitudes. The advantage of LCTMag at high latitudes is highlighted in \cref{tab:lorentzian_fits}. The table compares the Lorentzian fit parameters determined from the 1D line profiles of representative inertial modes obtained from LCTMag, LCTGran and RDA. We see that LCTMag has a much higher S/N when compared to RDA at higher latitudes ($m=1$). At lower latitudes, however, we see that RDA outperforms the LCTMag ($m=2,8$), albeit by a small margin. LCTGran has a lower S/N, compared to the other methods for all example modes. For the HFR modes ($m=13$), the excess power is relatively weak in these latitudinally averaged line profiles, as the modes are more clearly isolated in the $l=m+1$ spherical harmonic component.

\section{Extracting inertial mode latitudinal eigenfunctions \label{sec:ef_extract}}

\subsection{Singular value decomposition method}

Once a mode is identified at a particular value of $m$, we can filter the Fourier transform of the horizontal velocity $\mathbf{U}^{\pm}(\omega, \theta, m)$ around the mode frequency, $\omega_0$, and transform back to the time domain to obtain the filtered flow velocity associated with the mode,
\begin{equation}
    \mathbf{u}^\pm_0 (t,\theta) = \sum_\omega e^{\textrm{i} \omega t} 
    F(\omega-\omega_0) \mathbf{U}^{\pm}(\omega, \theta, m) .
\end{equation}
The filter $F(\omega - \omega_0)$ is a flat-top window with raised-cosine flanks centred on the mode frequency, $\omega_0$, with a total passband width of approximately two to four times the mode linewidth.

We followed the singular value decomposition (SVD) approach used by \cite{Proxauf2020} to extract the latitudinal dependence of the mode shapes. The SVD may be written as
\begin{equation}
    \mathbf{u}_0^\pm (t,\theta) = \sum_{q=1}^{2 n_\theta} \sigma_q T_q(t) \mathbf{X}^*_q(\theta) ,
\end{equation}
where $\sigma_1 \ge \sigma_2 \ge \cdots \ge \sigma_{2n_\theta} \ge 0 $ are the singular values, $n_\theta$ is the number of points in latitude, and $T_q(t)$ and $\mathbf{X}_q(\theta)$ are the singular vectors in time and space respectively. The left singular vectors are orthonormal by construction, that is $\sum_t |T_q(t)|^2 =1$ for each $q\ge 1$.

We assume that the time dependence of the mode is well approximated by the singular vector associated with the largest singular value, $T_1(t)$. To estimate the latitudinal dependence of the velocity eigenfunction of the mode, $\mathbf{v}_0^{\pm}(\theta)$,  we project the full $\bu^\pm_0(t,\theta)$ onto $T_1(t)$ to obtain
\begin{equation}
    \mathbf{v}_0^{\pm}(\theta) =  \sum_t  T_1^*(t)\  \mathbf{u}_0^{\pm}(t, \theta) ,
\end{equation}
which is a complex vector. This approach yields a robust estimate of the eigenfunction even at high latitudes, where the data coverage is most incomplete. 

\begin{figure*}
\centering
\begin{adjustbox}{max width=\textwidth, max totalheight=0.9\textheight, keepaspectratio}
\begin{minipage}{\textwidth}
    \centering
    \includegraphics[width=\textwidth]{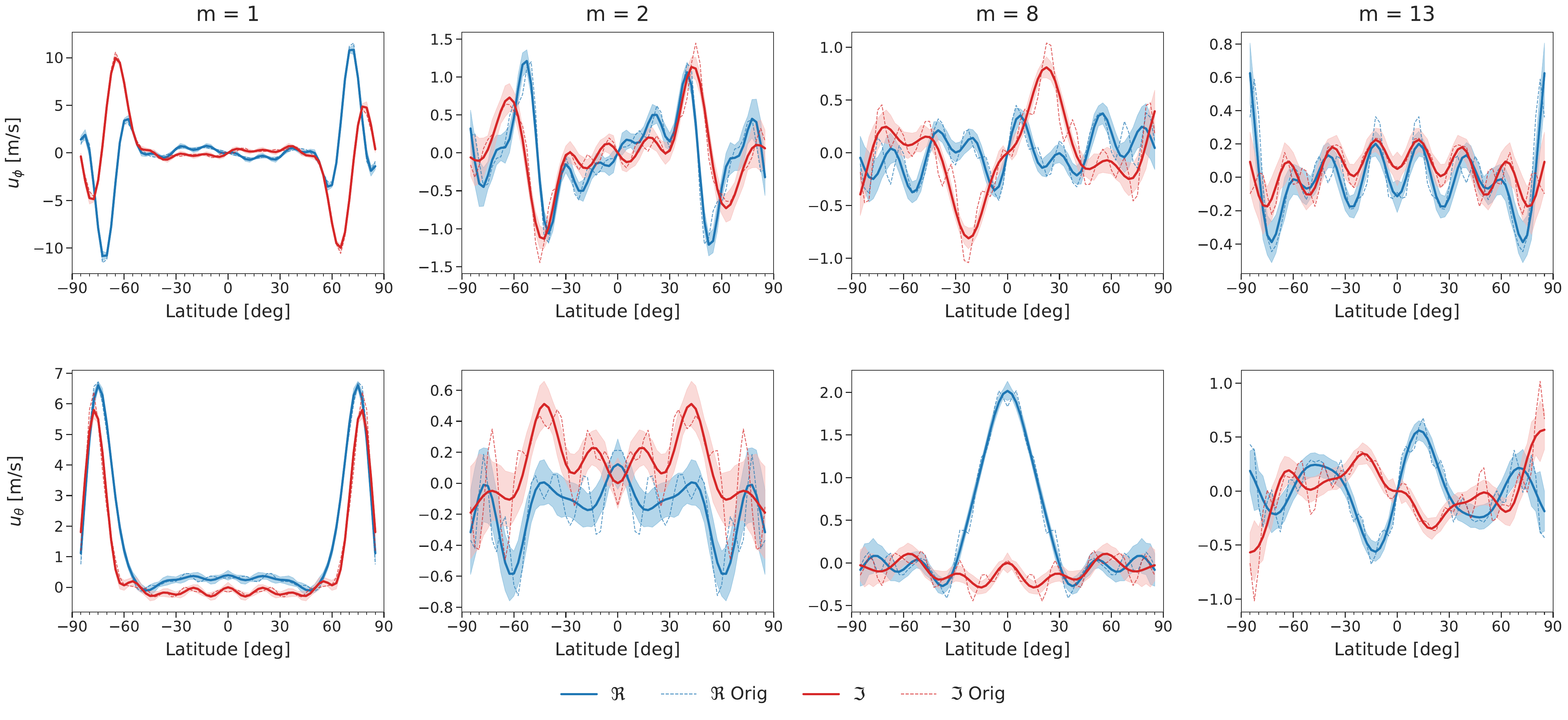}
    
    \includegraphics[width=\textwidth]{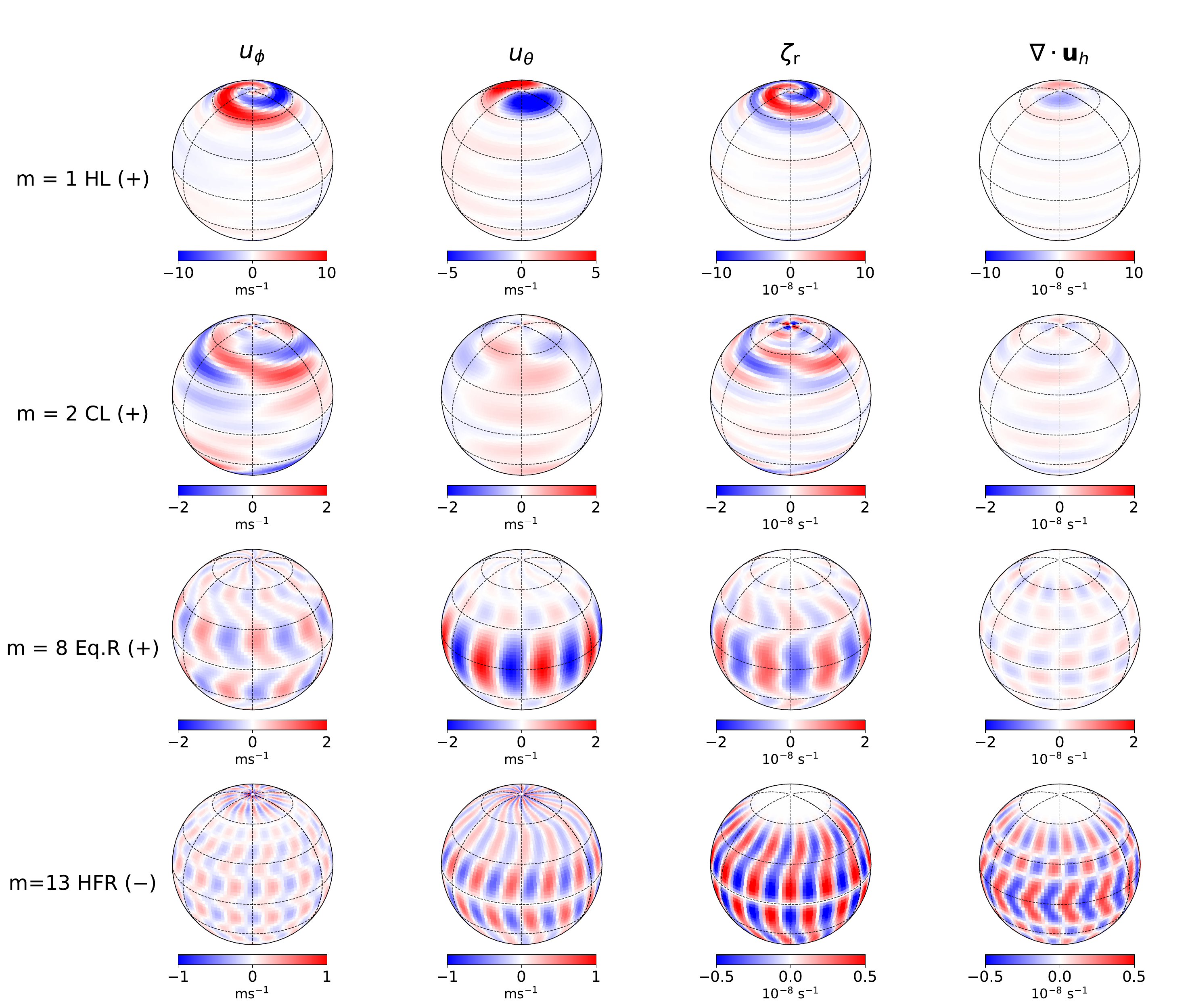}
\end{minipage}
\end{adjustbox}

    \caption{\label{fig:2d_efs}
    \textit{Top}: 1D eigenfunctions associated with selected inertial modes obtained from SVD. Solid curves show the filtered eigenfunctions, while dashed curves show the original unfiltered eigenfunctions. The shaded regions indicate the 1$\sigma$ uncertainty estimated from Monte Carlo realisations of the discarded high-$l$ coefficients. The filtering procedure is described in \cref{sec:ef_filter}.
    \textit{Bottom}: Corresponding 2D eigenfunctions derived from the LCTMag flow maps. Columns show $u_\phi$, $u_\theta$, $\zeta_\mathrm{r}$, and $\nabla \cdot \mathbf{u}_h$. The $m$ value is indicated on the left; $(+)$ and $(-)$ denote modes that are symmetric and antisymmetric, respectively, in $u_\theta$.
}
\end{figure*}

\begin{figure*}
  \resizebox{\hsize}{!}{\includegraphics{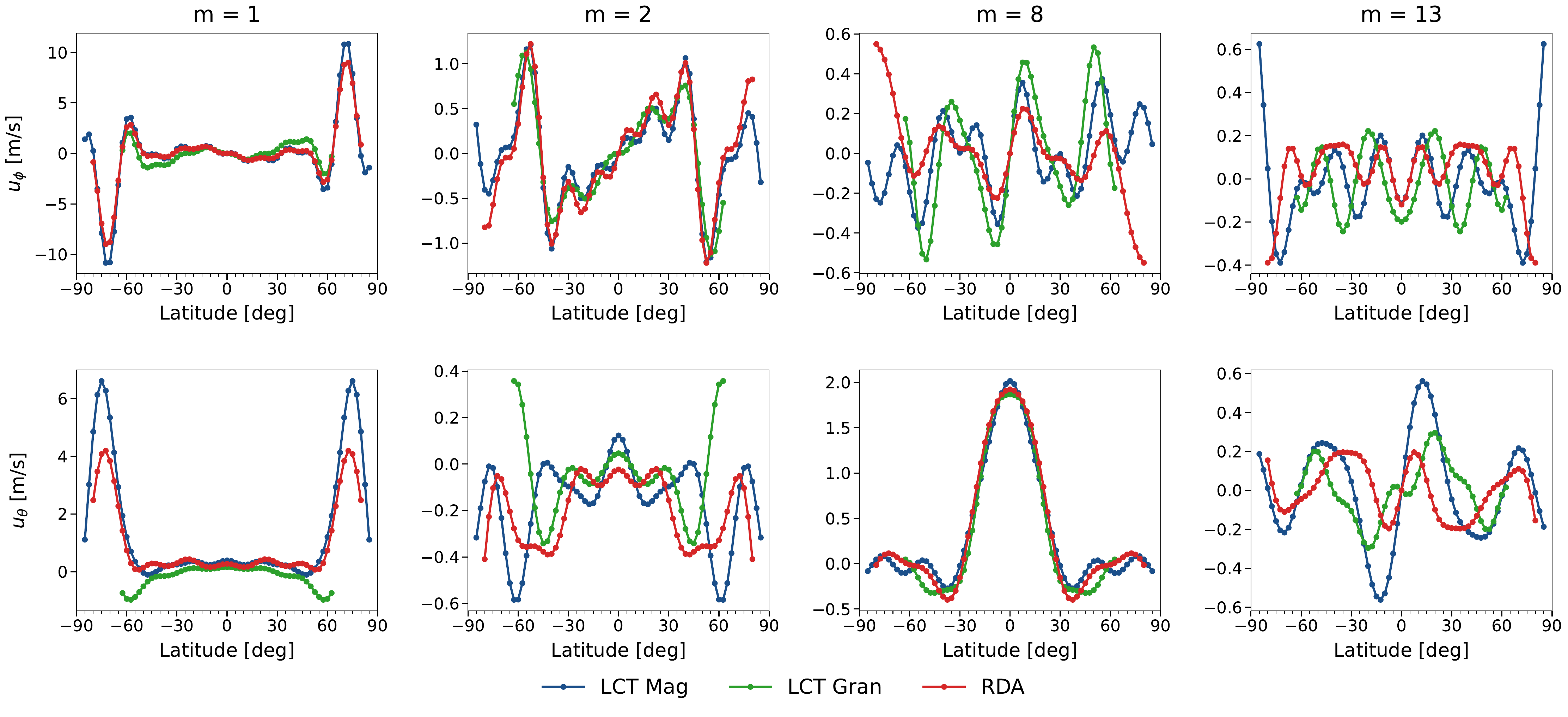}}
  \caption{  \label{fig:1d_ef_compare}Comparison of the real parts of the 1D eigenfunctions associated with the selected inertial modes obtained from LCTMag (blue), LCTGran (green), and RDA (red). The eigenfunctions have been smoothed and phase-aligned. The LCTGran eigenfunctions are limited to the extent of validity (see Fig.~\ref{fig:ccf_widths_compare}.)
  }
\end{figure*}

\subsection{Noise in the eigenfunction \label{sec:ef_filter}}

Although SVD removes the major uncorrelated noise components, the eigenfunctions still exhibit rapid latitudinal oscillations as a function of latitude (see the dashed curves in the top panels of \cref{fig:2d_efs}). Given the small azimuthal wavenumbers, we assume these oscillations are due to residual noise. The residual noise is thought to arise from several sources, including aliasing of high-spatial-frequency processes such as convection and supergranulation, as well as signal-correlated noise. To suppress this residual noise, we decomposed the 1D eigenfunctions onto a basis of Legendre polynomials and obtain the coefficients associated with each $l$ via
\begin{equation}
f_{l,j} = \sqrt{\frac{2l+1}{2}}\int_{0}^{\pi} v^{\pm}_j(\theta)\, P_l(\cos\theta)\, \sin\theta \, d\theta,
\end{equation}
where $v^{\pm}_j(\theta)$ is the latitudinal eigenfunction for the flow component, $j$, and $P_l(\cos\theta)$ are the Legendre polynomials of degree, $l$. The coefficients $f_{l,j}$ thus quantify the contribution of each harmonic degree to the eigenfunction. We then consider the power of the coefficients $|f_{l,j}|^2$ and plot it as a function of $l$. \Cref{fig:fl_power} shows the example of the $m=1$ symmetric HL mode. To determine a threshold, the 90\% confidence level is computed assuming $\chi^2$-distributed noise power with 2 degrees of freedom. We retain the $l$ values whose power exceeds the 90\% confidence level and that together account for 99\% of cumulative power. For noisier modes, where few coefficients exceed the confidence threshold, we instead adopt $l<15$ as the range contributing to the mode power. The discarded $l$ values are used to estimate the noise level. Our method involves randomising the phase for the discarded $l$ coefficients using Monte Carlo simulations. In the next step, we recomputed different realisations of the noisy eigenfunction. The error at each latitude is given by the standard deviation across different realisations. \Cref{fig:2d_efs} (top) shows the 1D eigenfunctions of the selected modes obtained from the LCTMag data before and after filtering in Legendre space, together with their associated uncertainties.

\section{2D Mode eigenfunctions and comparison with RDA \label{sec:rda_compare}}

For better visualisation, the 1D latitudinal eigenfunction can be converted to a 2D eigenfunction on the solar surface by 
\begin{equation}
    v^{\pm}_j(\theta, \phi) = 2\operatorname{Re}\left(v^{\pm}_j(\theta) e^{-\textrm{i}m\phi}\right). 
\end{equation}
The eigenfunctions for radial vorticity $\zeta_\mathrm{r}$ and horizontal divergence $\nabla \cdot \mathbf{u}_h$ are computed analogously. \Cref{fig:2d_efs} (bottom) shows the 2D eigenfunctions on the solar surface for the selected modes from LCTMag. We observe that the horizontal divergence is significantly smaller than the radial vorticity, except for the HFR mode. This implies the existence of radial motions associated with the HFR mode, in agreement with the findings of \cite{Hanson2022}. The horizontal divergence of the HFR mode is symmetric and resembles the spherical harmonics $Y^m_m$, consistent with the HFR modes being most prominent in the $l=m$ component of $\nabla \cdot \mathbf{u}_h$. 

The 1D eigenfunctions are better suited for comparing the different flow measurement techniques. \Cref{fig:1d_ef_compare} compares the real parts of the 1D eigenfunctions for the selected modes obtained using the three flow measurement techniques. The eigenfunctions from all three methods are extracted and denoised using the same procedure described in \cref{sec:ef_extract}. They are then phase-rotated to align the phase at the latitude of maximum amplitude. LCTGran becomes unreliable above approximately $\pm 65\degree$ latitude. LCTMag extends to higher latitudes than both LCTGran and RDA, reaching approximately $85\degree$, compared with $80\degree$ for RDA. For clarity, the unreliable high-latitude regions of the LCTGran and RDA eigenfunctions are not shown. We find reasonable agreement among the three methods for all example modes. The differences between the methods are within the error bars in most cases. For the case of the $m=13$ HFR mode, the agreement is only marginal because the signal is very weak in all three methods.

\section{Newly identified mode at $m=3$ \label{sec:m3_mode}} 

\begin{figure*}
\centering
\includegraphics[width=0.9\hsize]{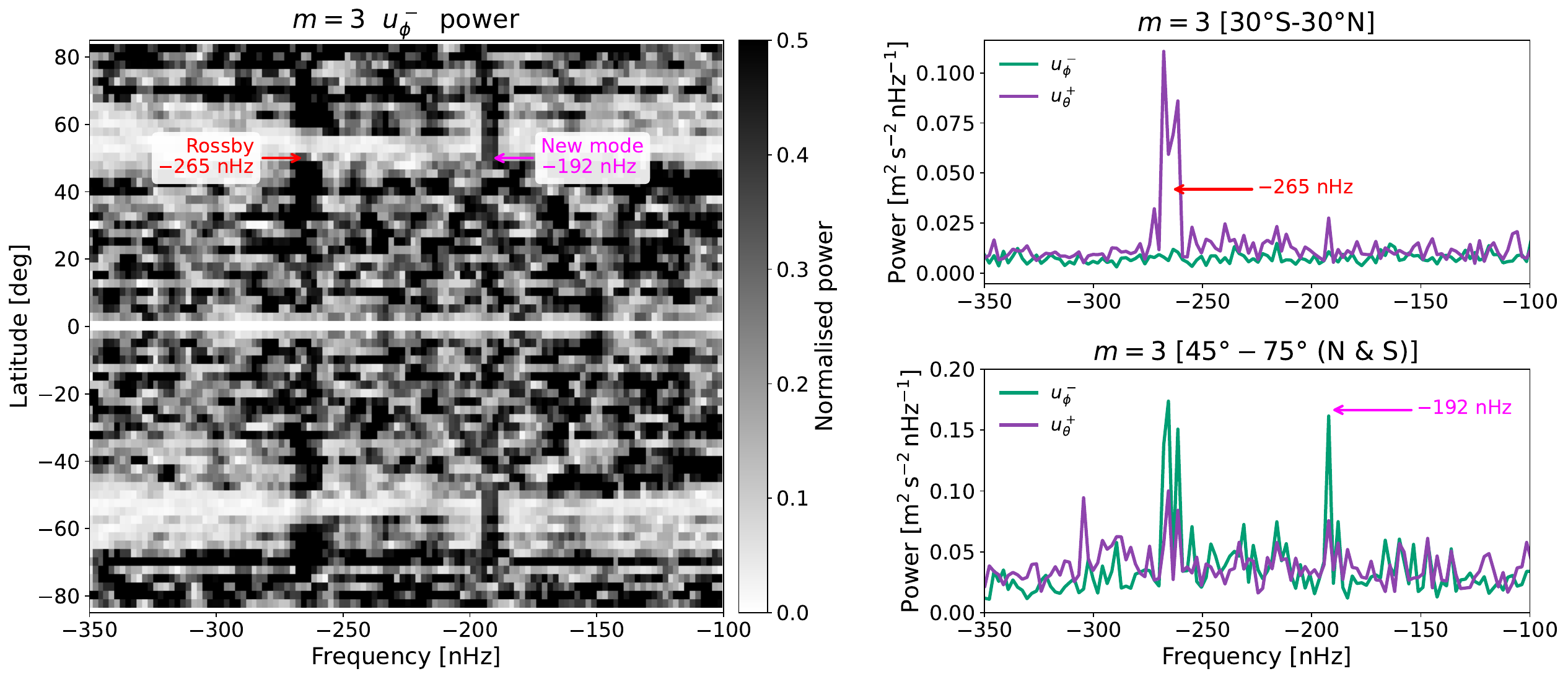}
\caption{  \label{fig:m3_rossby_hl}PSD of $u_\phi^-$ showing the newly identified $m=3$ mode. \textit{Left}: PSD as a function of latitude and frequency. The power at each latitude is normalised to the peak near $-192$~nHz to emphasise the new mode. The 2D power spectrum has been smoothed along the frequency axis using a running average with a bin width of 6.6~nHz. \textit{Right}: Latitude-averaged 1D PSD profiles for $u_\phi^-$ and $u_\theta^+$ in the latitude bands indicated in the panel titles. Arrows indicate the central frequencies of the Rossby mode ($-265$~nHz) and the new mode ($-192$~nHz). The Rossby mode (red arrow) shows power across both equatorial and mid-to-high latitudes, whereas the new mode (magenta arrow) has most of its power confined at higher latitudes.
}
\end{figure*}

\begin{figure*}
\centering
\includegraphics[width=0.9\hsize]{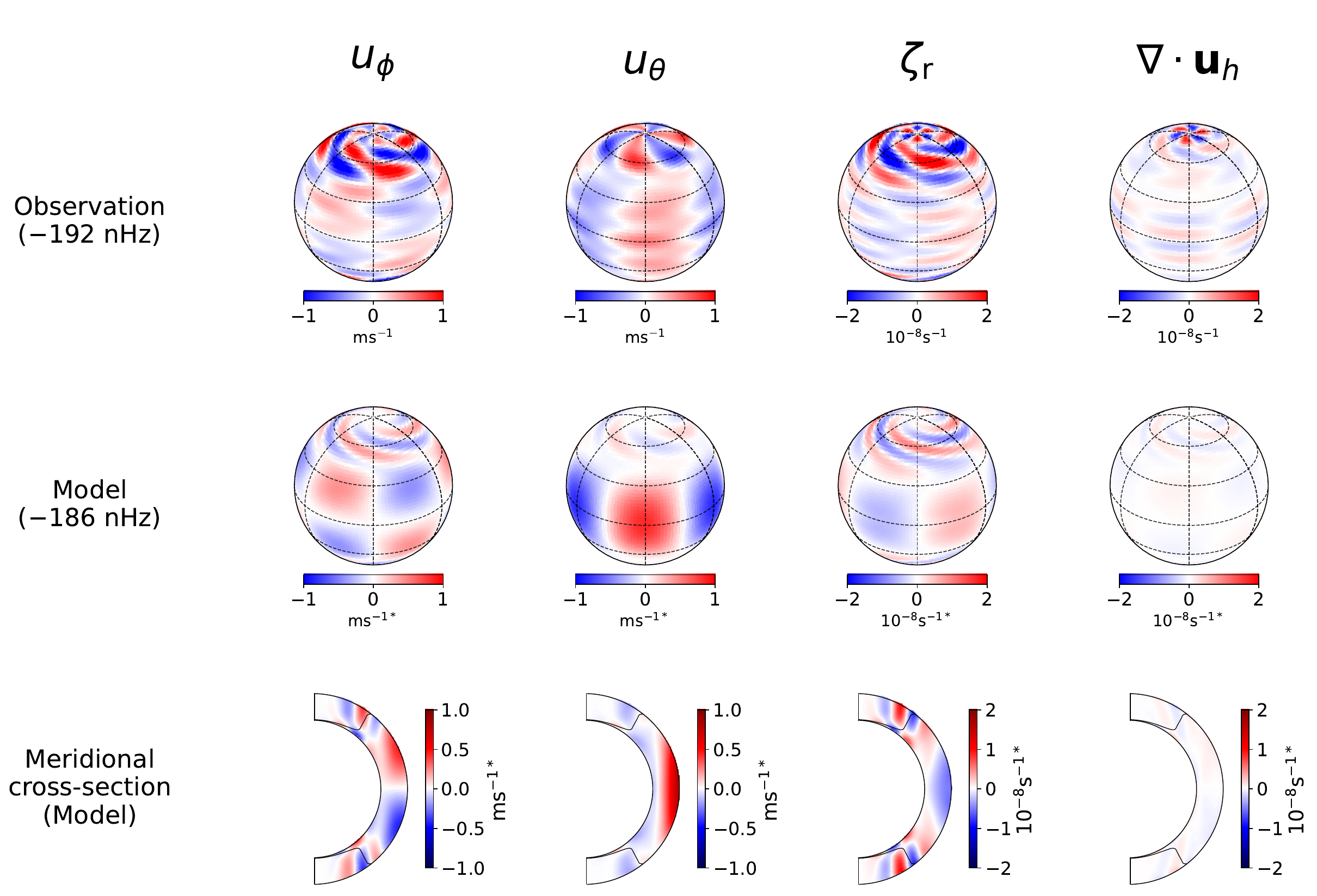}
\caption{  \label{fig:m3_new}2D eigenfunctions for the new $m=3$ mode along with those of the closest $m=3$ mode found in the linear model of \cite{sup2024}. The latter has previously been identified as an $n=1$ (one radial node) mode \citep[e.g.,][]{bekki2022,jain2024}. We also plot the meridional cross sections for this mode, with the black solid lines indicating the critical layers. The model eigenfunctions are expressed in physical units (denoted by an asterisk) by multiplying a dimensional factor. The two velocity components in the model are scaled such that $u_\theta$ at the equator takes approximately the same value in the model and the observations.  
}
\end{figure*}

An additional advantage of LCTMag is that it provides an independent means of identifying new modes. In our analysis, we identified a prominent peak at $-192$~nHz in the $u_\phi^-$ power spectrum that had not been identified previously. In  \cref{fig:m3_rossby_hl}, we show the 2D power spectrum at $m=3$ alongside the latitudinally averaged 1D line profiles. The peak exhibits enhanced power at high latitudes and is well separated from the nearby peak of the classical $m=3$ equatorial Rossby mode at $-265$~nHz. The top panels in \cref{fig:m3_new} show the 2D eigenfunctions of this new mode. We observe a spiral pattern at high latitudes in both $u_\phi$ and $u_\theta$, with a significant velocity pattern at low latitudes. The $m=3$ Rossby mode at $-265$~nHz exhibits this behaviour as well, having a high latitudinal spiral pattern in $u_\phi$ above $60\degree$, while simultaneously displaying a significant cell-like velocity pattern in the $u_\theta$ eigenfunction \citep{Gizon2021}.

With respect to the literature, \cite{hathaway2021} performed a spherical harmonic decomposition of the high-latitude flows and found that the high-latitude signal is dominated by an $m=1$ component at a frequency of approximately $-80$~nHz, with weaker $m=2$ and $m=3$ components appearing at double and triple that frequency (these are harmonics due to the fact that inertial mode eigenfunctions are not spherical harmonics). The value of $-80$~nHz is close to the mode frequency of $-86.3$~nHz reported by \cite{Gizon2021} for the $m=1$ high-latitude mode ($-87.2$~nHz in the present paper). Our new mode at $-192$~nHz is clearly distinct from the third harmonic of the $m=1$ high-latitude mode, as they are separated by about $70$~nHz. It also exhibits a spatial eigenfunction that differs markedly from the high-latitude spiral pattern reported by \cite{hathaway2021}. Therefore, there should not be any confusion between these signals.

Despite observing the eigenfunction, studying the actual mode behaviour rests on our being able to correctly identify the mode in numerical models. This is a major hurdle in inertial-mode helioseismology, as mode identification proves to be very tricky. A possible identification is provided by the closest mode found in the compressible linear model of \cite{sup2024}, namely an $n=1$ Rossby mode with one radial node. Alongside the eigenfunctions for the new $m=3$ mode in LCTMag, \cref{fig:m3_new} shows the corresponding model eigenfunctions for $u_\phi$, $u_\theta$, $\zeta_\mathrm{r}$ and $\nabla \cdot \mathbf{u}_h$. The model exhibits a spiral pattern at high latitudes close to the critical layers. The frequencies differ by 6~nHz. The eigenfunctions also show differences with the model, displaying high velocities at low latitudes. However, these differences could come from the incomplete knowledge of the underlying physics. For example, the spiral pattern and horizontal divergence change with changing turbulent viscosity \citep{bekki2022}, indicating that a number of gaps remain in the overall understanding of inertial mode physics that would require further study.

\section{Conclusions and discussion \label{sec:conclusions}}

In this work, we applied magnetic-feature tracking to study the solar inertial modes and derived their horizontal velocity eigenfunctions from 14 years of HMI observations. When applied to HMI data, LCTMag provides an independently measured horizontal flow dataset from a non-helioseismic technique, which offers various advantages such as more extensive spatial coverage at higher latitudes and leads to better S/N. We also show a comparison of the size of the average feature tracked using LCTGran and LCTMag (see \cref{fig:ccf_widths_compare}). This provides an estimate of the latitude range for which each technique yields reliable results. For LCTGran, we see that the reliability of the flow measurements drops dramatically at latitudes higher than $65\degree$. This is caused by the reduced apparent resolution of granules owing to foreshortening at large viewing angles. In contrast, we find that LCTMag tracks magnetic features with sizes ranging around 2--5~Mm up to $\pm85\degree$, with gradients in the north-south and east-west direction also arising from foreshortening effects.

We find that the background noise is significantly reduced by 80\% in the power spectrum of inertial modes using LCTMag when compared to LCTGran (see \cref{fig:pow2d_rossby} and \cref{fig:pow1d}). This leads to a clearer identification of the Rossby mode ridge in the power spectrum. This is even more evident when we look at the high-latitude modes. Due to the lower convection noise, we can also see the HFR mode ridge directly in the $u_\theta$ component of horizontal flow. 

We also estimated the random errors of the latitudinal eigenfunctions using the discarded high-$l$ coefficients from the Legendre polynomial decomposition. The value of this error depends on the particular mode, with a typical value that is close to $\sim$0.2--0.4~m/s. A comparison between different flow measurement methods also reveals a clear consistency with minor deviations within error bars. 

The LCTMag dataset can also help us find new inertial modes, such as the $m=3$ mode described in \cref{fig:m3_new}. This mode has a similar eigenfunction to the $m=3$ Rossby mode, which exhibits characteristics of a typical Rossby mode, but with a spiral pattern similar to HL modes. This new mode has a frequency of $-192$~nHz, as compared to $-265$~nHz for the classical $m=3$ equatorial Rossby mode. Mode identification remains a major challenge because linear models do not provide information about mode amplitudes. We highlight this using a comparison with the closest $m=3$ mode computed using a linear compressible model. This new dataset provides improved S/N and spatial coverage, making it useful for confronting the models and improving our understanding of the solar convection zone.

\begin{acknowledgements}
Author contributions: NJ, LG, Z-CL designed research, NJ wrote the LCT code and produced the LCT flow maps, RSB and YM provided the RDA flow maps, NJ, Z-CL and DF analysed the observations, SM identified the new $m=3$ mode in the theoretical spectrum (eigenvalue solver). 

NJ produced the first draft of the paper, and all authors contributed to the final manuscript. We would like to thank A.C.~Birch, R.H.~Cameron, J.~Schou and D.~Yang for helpful discussions and comments. The HMI data are courtesy of NASA/SDO and the HMI Science Team. The eigenfunction data generated in this work have been deposited on Edmond, the Open Research Data Repository of the Max Planck Society, and are publicly available at \url{https://doi.org/10.17617/3.37W3ER}. We acknowledge partial financial support from the ERC Synergy Grant WHOLESUN (grant agreement No. 810218) to LG. NJ is a member of the International Max-Planck Research School for Solar System Science at the University of Göttingen. 
\end{acknowledgements}

\bibliographystyle{aa}
\bibliography{ref}

\begin{appendix}
\onecolumn
\section{Additional figures}

\begin{figure*}[h!]
\sidecaption
\includegraphics[width=12cm]{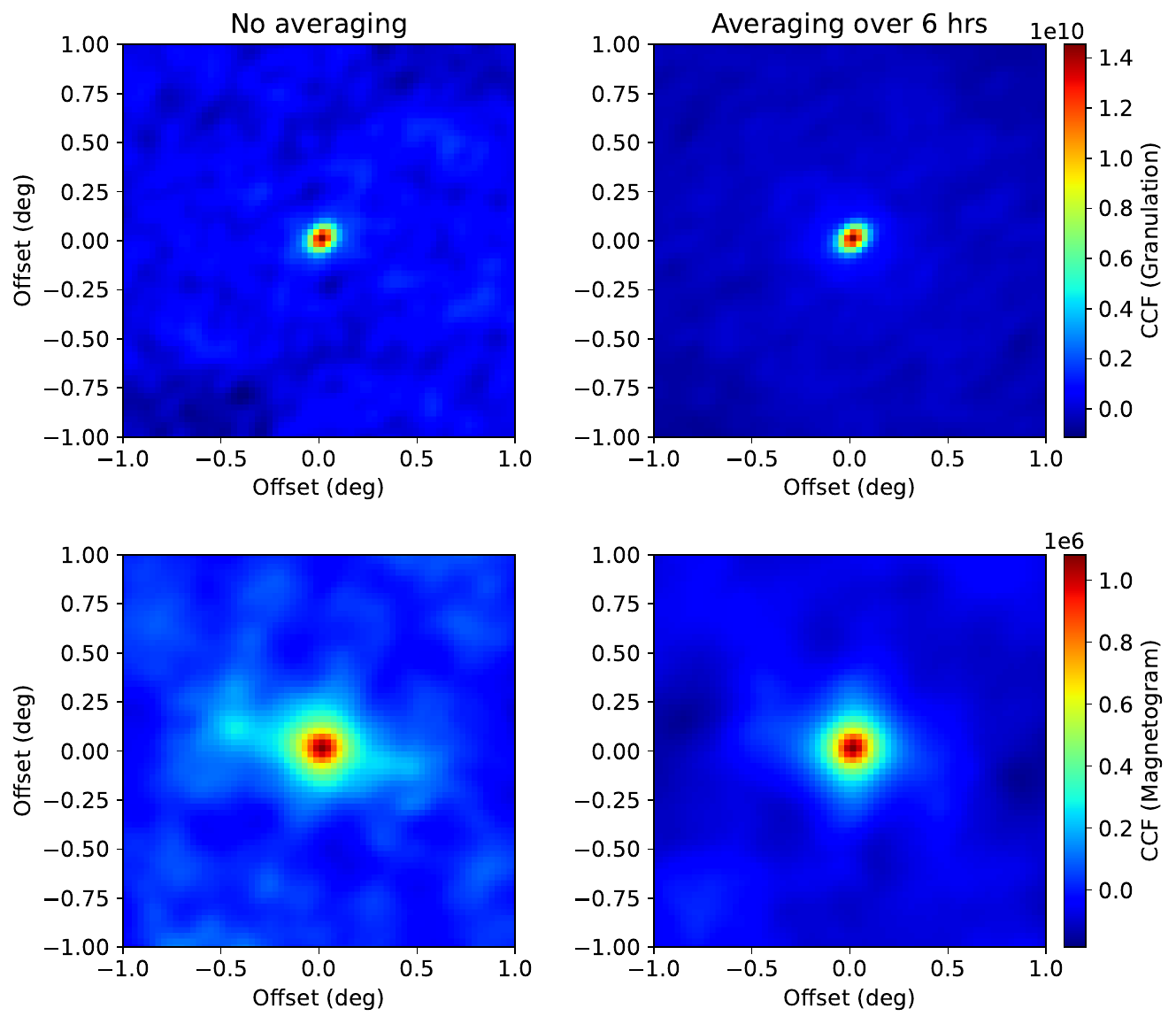}
  \caption{  \label{fig:ccf_2d}Example cross-correlation functions computed from continuum intensity images (\textit{top}) and magnetograms (\textit{bottom}), with and without temporal averaging, for a target location at Stonyhurst heliographic coordinates $(\theta, \phi) = (60\degree, 30\degree)$. The plots are restricted to a $2\degree\times2\degree$ box for clarity.}
\end{figure*}

\begin{figure*}[h!]
\sidecaption
\includegraphics[width=12cm]{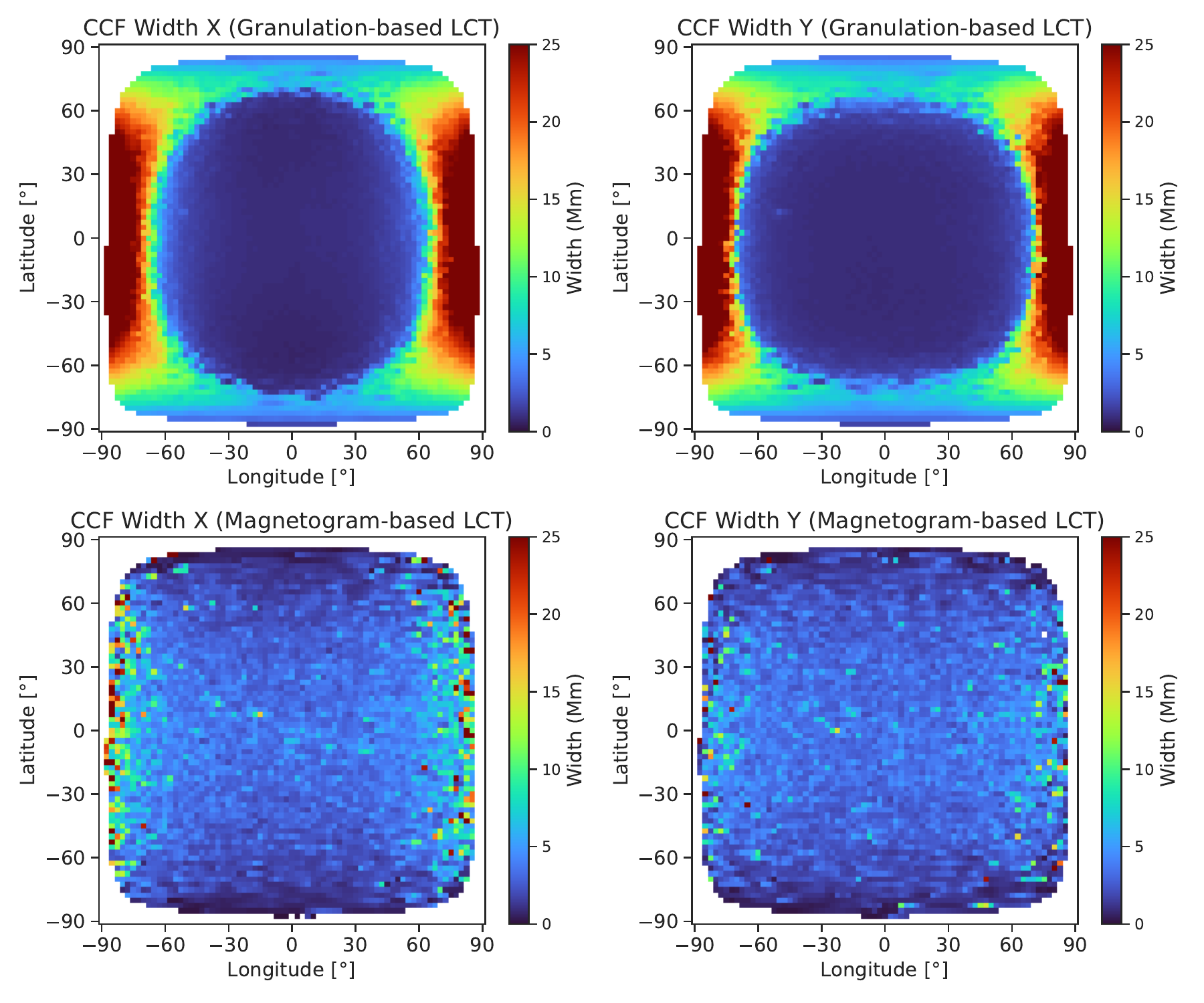}
  \caption{\label{fig:ccf_widths_compare}FWHM of the CCFs from LCTGran (\textit{top panels}) and LCTMag (\textit{bottom panels}), measured in local Cartesian coordinates. The CCF widths provide a measure of the characteristic size of the tracked features. The analysis uses data from 2018-12-15 00:00--06:00 TAI, when the $B_0$ angle was nearly zero and solar activity was very low. 
  }
\end{figure*}

\begin{figure*}
\sidecaption
    \includegraphics[width=12cm]{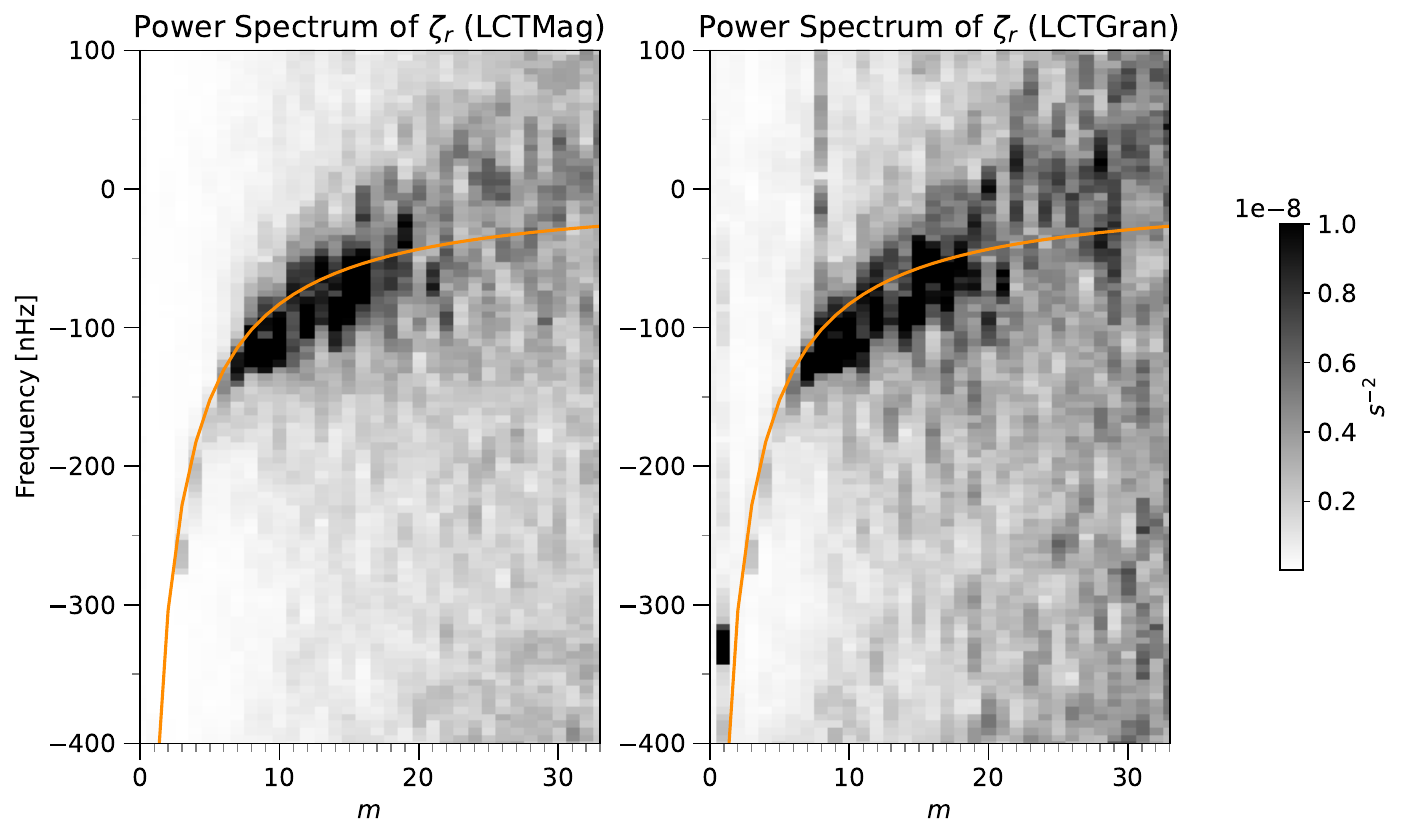}
  \caption{  \label{fig:pow2d_rvort_mag_gran}Radial vorticity ($\zeta_\mathrm{r}$) power spectra from LCTMag (\textit{left}) and LCTGran (\textit{right}) at different $m$ for the spherical harmonics $l=m$. The orange line corresponds to the theoretical dispersion relation of the sectoral Rossby modes. The power spectra are binned down to 5~nHz and convolved with a window of 25~nHz to suppress the noise.}
\end{figure*}

\begin{figure*}
\sidecaption
    \includegraphics[width=12cm]{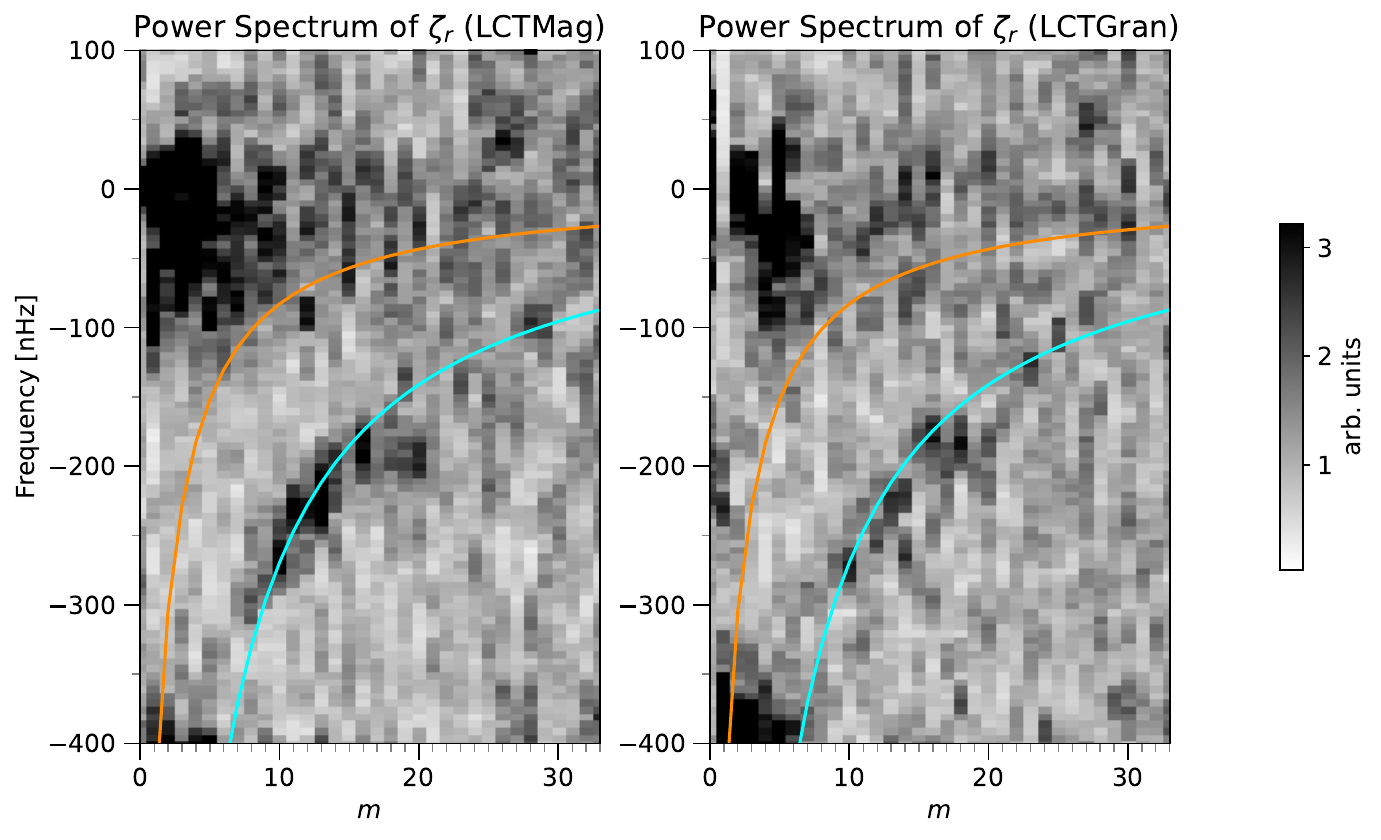}
  \caption{  \label{fig:pow2d_rvort_hfr}Normalised radial vorticity ($\zeta_\mathrm{r}$) power spectra from LCTMag (\textit{left}) and LCTGran (\textit{right}) at different $m$ for the spherical harmonics $l=m+1$. The orange line corresponds to the theoretical dispersion relation of the sectoral Rossby modes, while the cyan line is roughly three times the sectoral Rossby mode frequency. Each power spectra are binned down to 5~nHz and convolved with a window of 25~nHz to suppress the noise. The colour scale is saturated to 60\% of the maximum power at $m=13$ in the left panel; the same colour scale is used unchanged in the right panel.}
\end{figure*}

\begin{figure*}
\sidecaption
    \includegraphics[width=12cm]{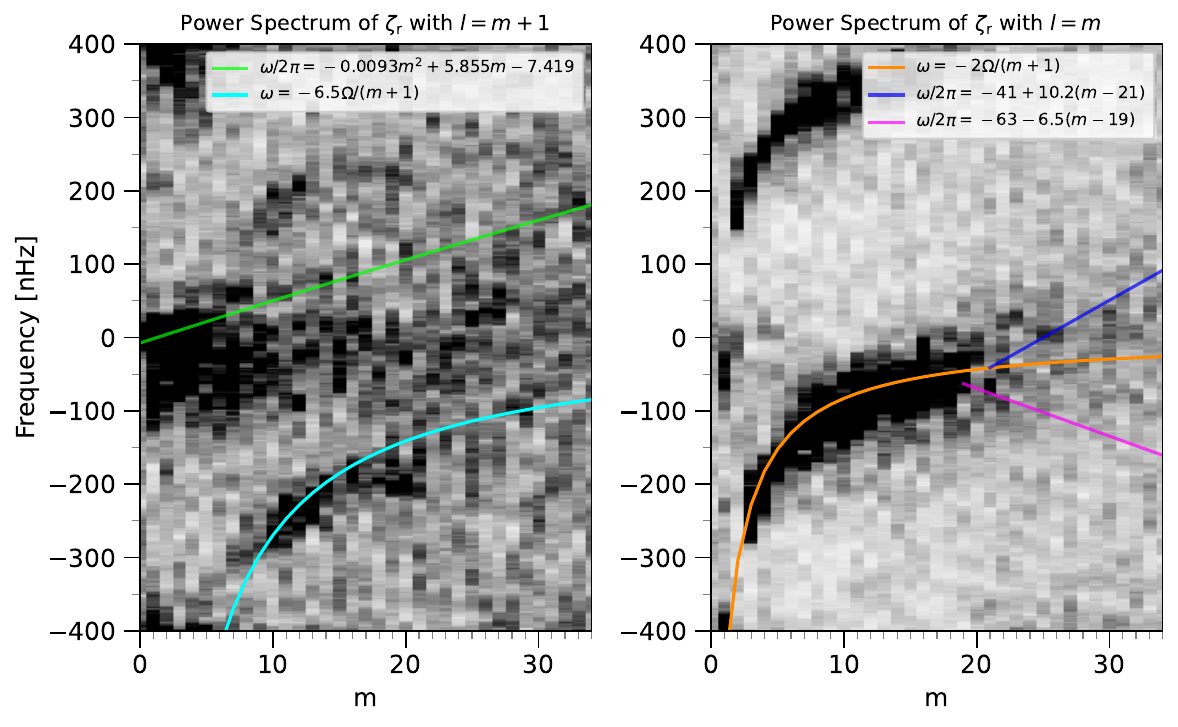}
  \caption{  \label{fig:pow2d_hanasoge}
  Normalised radial vorticity ($\zeta_\mathrm{r}$) power spectra from LCTMag using the $l=m+1$ (\textit{left}) and $l=m$ (\textit{right}) components. The orange line in the right panel shows the theoretical dispersion relation of the classical sectoral Rossby modes. The cyan line in the left panel shows an empirical guide, $\omega=-6.5\Omega_\text{Carr}/(m+1)$, chosen to approximately follow the HFR ridge reported by \cite{Hanson2022}. The green line shows the quadratic fit to the $R_1$ ridge presented in Fig.~1 of \cite{hanasoge2026}, while the blue and magenta lines in the right panel show the straight-line fits to the two ridges presented in \cite{hanasoge2026evidence}. In the latter two cases (green, blue, and magenta lines), no significant excess power is detected by LCTMag along the reported ridges.}
\end{figure*}

\begin{figure*}
\sidecaption
  \includegraphics[width=12cm]{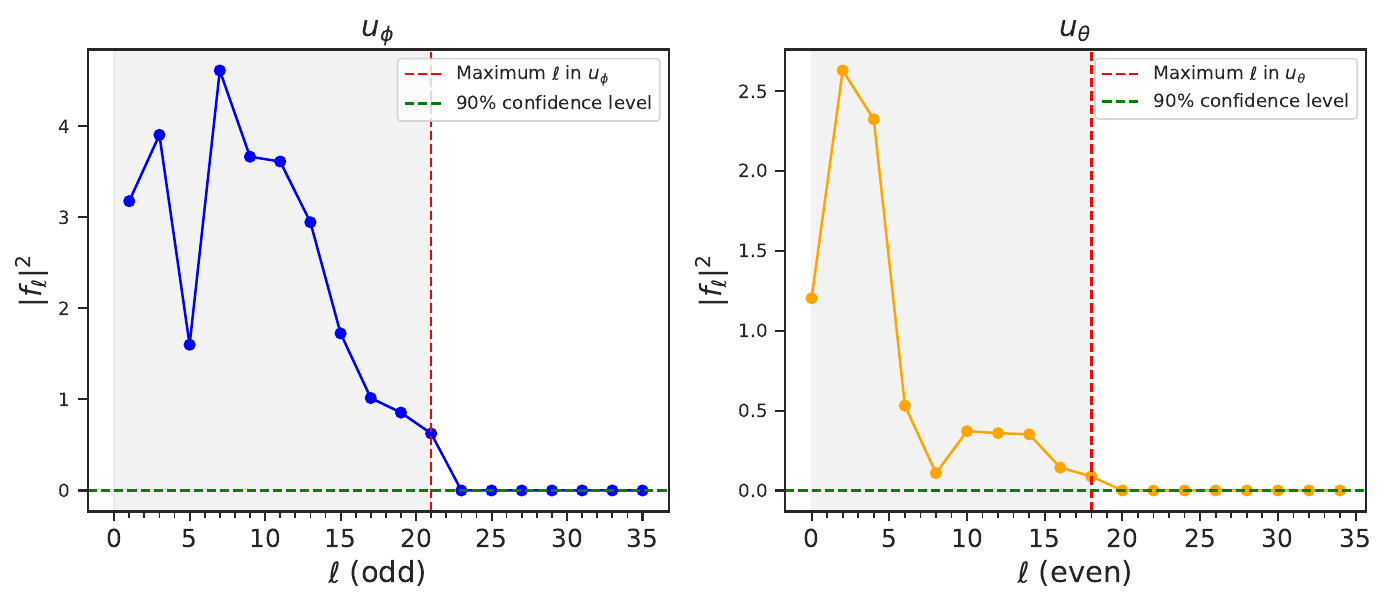}
  \caption{  \label{fig:fl_power}Power spectrum of the Legendre polynomial coefficients for the $m=1$ high-latitude mode. The 1D eigenfunction of the mode is decomposed into a basis of Legendre polynomials, and the power is plotted as a function of $l$. The dashed line represents the 90\% confidence level from the assumed $\chi^2$ distribution (see \cref{sec:ef_filter}). Since the $m=1$ mode is symmetric in $u_\theta$, the $u_\theta$ and $u_\phi$ components contain only even and odd values of $l$, respectively. Here, $l$ refers to the degree of the Legendre polynomial $P_l(\cos\theta)$, not the associated Legendre polynomial $P_l^m$ used in spherical harmonics.
  }
\end{figure*}

\end{appendix}

\end{document}